\documentclass[aps, superscript address,float fix,one column,reprint,footinbib]{revtex4-1} 

\usepackage{amsmath,amssymb,bm,epsfig}
\usepackage{color}
\usepackage{natbib}
\usepackage{hyperref} 
\usepackage{ulem}
\usepackage{graphicx}
\usepackage{wasysym}
\RequirePackage{lineno}

\usepackage[utf8]{inputenc}

\usepackage{xcolor,colortbl}
\usepackage{pifont}
\usepackage{footnote}

                \def\lsim{\mathrel{\rlap{\lower4pt\hbox{\hskip1pt$\sim$}}
    \raise1pt\hbox{$<$}}}                \def\gsim{\mathrel{\rlap{\lower4pt\hbox{\hskip1pt$\sim$}}
    \raise1pt\hbox{$>$}}}                
\newcommand{\as}{\alpha_\mathrm{s}}

\newcommand{\kt}{k_{t}}

\newcommand{\btv}{\textbf{b}}
\newcommand{\rtv}{\textbf{r}}
\newcommand{\ktv}{\textbf{k}}
\newcommand{\ptv}{\textbf{p}}

\def\Nch{N_{\rm ch}^{\rm rec}}

\def\Ntrig{N_{\rm trig}}

\def\as{\alpha_S}
\def\sp{S_\perp}
\def\qs{Q_{\!_S}}
\def\qnot{Q_{\!0}}

\def\nc{N_c}
\def\cf{C_F}
 
\def\qp{ {\bf q}_{_T} } 
 
\def\pp{ {\bf p}_{_T} } 
\def\kp{ {\bf k}_{_T} } 
\def\xt{ {\bf x}_{_\perp} } 
 
\newcommand{\kpn}[1]{ {\bf k}_{#1\perp} } 
\newcommand{\qpn}[1]{ {\bf q}_{#1\perp} }
\newcommand{\ud}{\mathrm{d}}

\def \beq{\begin{equation}}
\def \eeq{\end{equation}}
\def \beqa{\begin{eqnarray}}
\def \eeqa{\end{eqnarray}}
\def\ra{\right>} 
\def\la{\left<} 
\newcommand{\rb}[1]{ {\bf r}_{#1} }

\newcommand{\tr}{\, \mathrm{Tr} \, }

\begin{document}

\title{Collectivity in small collision systems : an initial state perspective}

\author{S\"{o}ren Schlichting}
\affiliation{Physics Department, Brookhaven National Laboratory,
  Upton, NY 11973, USA
}
\affiliation{Department of Physics, University of Washington, Seattle, Washington 98195-1560, USA}

\author{Prithwish Tribedy}
\affiliation{Physics Department, Brookhaven National Laboratory,
  Upton, NY 11973, USA
}

\begin{abstract}
Measurements of multi-particle correlations in the collisions of small systems such as $p+p$, $p/d/^3He\!+\!A$ show striking similarity to the observations in heavy ion collisions. A number of observables measured in the high multiplicity events of these systems resemble features that are attributed to collectivity driven by hydrodynamics. However alternative explanations based on initial state dynamics are able to describe many characteristic features of these measurements. In this brief review we highlight some of the recent developments and outstanding issues in this direction. 

\end{abstract}

\maketitle

\section{Introduction}

Relativistic collisions of hadrons and nuclei at the modern colliders provide unique testing ground for QCD at high energies. Over a decade of experimental measurements for a wide range of energies and collision systems have been dedicated to study the properties of the matter formed in such collisions. A number of measurements have provided strong indications that the QCD matter formed in the collisions of two nuclei (A+A) behave like a strongly interacting fluid that exhibits collectivity~\cite{Arsene:2004fa, Back:2004je, Adams:2005dq, Adcox:2004mh}. Consistent measurements of strong radial and anisotropic flow, jet quenching etc. have convincingly established such properties of the medium. Small system collisions have initially been thought of providing benchmarking measurements for the observations in heavy ion collisions. 

Very recently several striking observations have been made in p+p, p+A, d+A and $^3He+A$ collisions~\cite{Khachatryan:2010gv, CMS:2012qk, Abelev:2012ola, Aad:2012gla, Adare:2014keg, Adamczyk:2015xjc, Adare:2015ctn, Aad:2015gqa, Khachatryan:2015lva}. Observations in high multiplicity events for such small collision systems seem to resemble features that are common to A+A collisions and very often attributed to decisive signatures of collectivity due to hydrodynamic evolution. Such observations include appearance of azimuthal correlations that extend in long-range rapidity known as ridge,  which is also quantified in terms of Fourier harmonic coefficient of $v_n$, strong multiplicity dependence of mean transverse momentum, HBT radii and many others.

An outstanding question remains whether such systematics have a collective origin that can be attributed to hydrodynamic evolution like in A+A collisions or a natural consequence due to initial state dynamics that appear in the final state observables or a combination of both. Several alternative approaches based on final state~\cite{Bozek:2013uha, Bozek:2013ska,Bzdak:2013zma,Qin:2013bha, Werner:2013ipa, Nagle:2013lja, Schenke:2014zha,Bzdak:2014dia,Kalaydzhyan:2015xba,Shuryak:2013ke,Ghosh:2014eqa} and initial state ~\cite{Kovner:2010xk,Kovner:2011pe,Levin:2011fb,Dusling:2012iga,Dusling:2012cg,Dusling:2012wy,Kovchegov:2012nd, Dusling:2013qoz, Kovchegov:2013ewa,Dumitru:2014dra,Dumitru:2014vka,Dumitru:2014yza,Gyulassy:2014cfa,Lappi:2015vha,Schenke:2015aqa,Lappi:2015vta,Dusling:2015rja} dynamics have provided competing arguments in this context. Interestingly, such a debate on the relative importance of initial state \cite{Wang:1991vx} and final state effects \cite{Levai:1991be} also took place in the early 90s while addressing the experimental data from $p+\bar{p}$ collisions at the Tevatron.

In this review we provide a brief overview on the experimental findings and outline a general theory perspective on the collective phenomena observed in small systems. Subsequently, we focus specifically on theoretical explanations based on the initial state dynamics. We discuss different theoretical approaches to the calculations and contrast the results with experimental findings. We conclude with a brief summary of the present status and perspectives for future studies.

\subsection{Experimental Overview}
The current debate of collectivity in small systems was triggered by the discovery of ridge like correlations that extend over a long range in rapidity in high multiplicity p+p collisions by the CMS collaboration~\cite{Khachatryan:2010gv}. Such long range structure of azimuthal correlations was previously seen in heavy ion collisions at RHIC~\cite{Adams:2005ph,Adare:2006nr,Alver:2008aa} and LHC~\cite{Chatrchyan:2012wg} and generally attributed to the nearly boost invariance structure of the azimuthal correlations driven by hydrodynamic flow. At the same time causality arguments suggest~\cite{Dumitru:2008wn} that such correlations must develop at the very early stages of the collisions, indicating the strong influence of initial state dynamics on such observable. While the observation of similar ridge like structures made in high multiplicity p+Pb collisions at the LHC was expected~\cite{CMS:2012qk, Abelev:2012ola, Aad:2012gla}, a surprise was that the signal strength at the same multiplicity was stronger in p+Pb compared to p+p. The latest inclusion to these measurements are the highest energy p+p collisions so far at 13 TeV~\cite{Aad:2015gqa, Khachatryan:2015lva}. Including the 13 TeV and 7 TeV data both ATLAS and CMS collaborations have demonstrated that the strength of ridge correlations (characterized by the near side yield) at a given multiplicity is independent of collision energy. However, a more pronounced energy dependence appears to be present in higher-order cumulants such as $c_{2}\{4\}=-v_{2}^{4}\{4\}$ which unlike at lower energies appears to exhibit a clear sign change as a function of multiplicity in 13 TeV $p+p$ collisions~\cite{CMS:2016yew}. Meanwhile RHIC, being a versatile machine, has collided $^3He+Au$ ~\cite{Adare:2015ctn} along with $d+Au$~\cite{Adare:2014keg, Adamczyk:2015xjc} and $p+Au$~\cite{Itaru2015qm} to perform similar measurements of azimuthal correlations. Such measurements indicate the presence of significant $v_2$ and $v_3$ in $p+Au, d+Au, ^3\!He+Au$, the systematics of which is very similar to what is commonly seen in $A+A$ collisions.  
\begin{figure*}[t]
\includegraphics[width=1\textwidth]{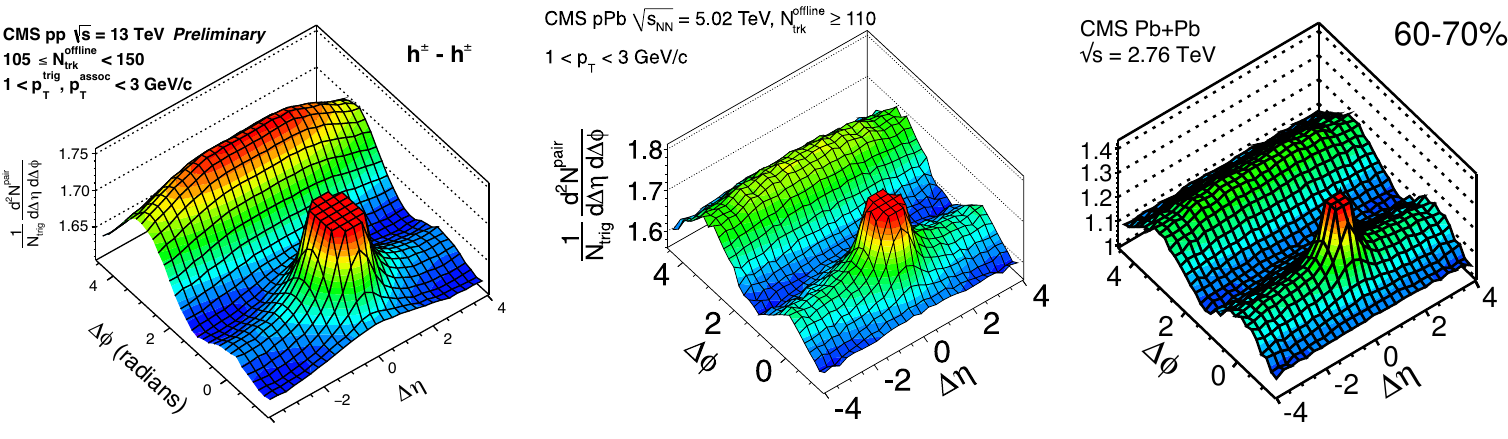}
\caption{\label{cms_ridge} Two-particle correlation function in relative pseudo-rapidity and azimuthal angle showing long range ridge-like structure in high multiplicity p+p, p+Pb as compared to peripheral Pb+Pb collisions. Figures are taken from \cite{Khachatryan:2015lva,CMS:2012qk, Chatrchyan:2012wg}}
\end{figure*} 
Analyses common to A+A have been repeated in both p+Pb and p+p at LHC by triggering on high multiplicity events. Measurements of higher order harmonics of azimuthal anisotropy $v_n$, mass dependence of $v_n$, higher order cumulants of $v_2\{n\}, n\ge4$ and many others are now available (for a comprehensive review we refer the reader to Ref~\cite{Dusling:2015gta, Loizides:2016tew}). 

By now a number of intriguing results observed in $p+p$ and $p/d/^3He+A$ have accumulated, including a strong multiplicity dependence of average transverse momentum $\la p_T\ra$ that is attributed to radial flow as well as the observation of sizable Fourier harmonic coefficients $v_n (p_T)$ up to $n=4$ and its higher order moments of the azimuthal correlation generally attributed to anisotropic flow. Most importantly, several characteristics, such as the mass dependence of both $\la p_T \ra$ and $v_{n}(p_T)$ have been found to be similar to what is seen in A+A collisions. 

However it is worth to mention that some striking contrasts also exist. Unlike in A+A collisions, where the observation of jet-quenching has been one of the pillars of the discovery of a strongly interacting Quark Gluon Plasma (QGP)~\cite{Chatrchyan:2011sx,Adler:2002tq}, so far no evidence of (mini) jet-quenching has been found in small systems~\cite{ATLAS:2014cpa,Adam:2014qja,Adare:2015gla,Adam:2016jfp}. Even though the standard jet-quenching analysis in small-systems is complicated due to trigger bias effects, the absence of such phenomena may provide important insights with regard to the theoretical interpretation of the observed phenomena.

\subsection{General theorectical perspectives}
It is useful to first address the question about the origin of long-range azimuthal correlations (shown in Fig.\ref{cms_ridge}) from a more general point of view and formulate our theoretical expectations based on previous observations in small and large systems. While causality arguments imply that any long range rapidity correlation must originate from the very early stages of the collision~\cite{Dumitru:2008wn}, this leaves open the question how the observed momentum space correlations are created dynamically during the space-time evolution. Specifically one can, at least from a theoretical point of view, distinguish two different mechanisms whereby momentum space correlations of hadrons produced in the final state reflect
\begin{itemize}
\item[i)] intrinsic momentum space correlations of the partons produced in initial (semi-) hard scatterings  
\item[and/or] 
\item[ii)]  position space correlations between initial state partons, e.g. the initial state geometry, which are transformed into momentum space correlations due to final state interactions. 
\end{itemize}

While in any realistic scenario, both kinds of correlations i) and ii) contribute to the long-range azimuthal correlations, their relative strength depends on the magnitude of final state effects. In low-multiplicity $p+p$ collisions  for example, the dominant source of long-range azimuthal correlations is due to the production of back to back (mini-) jets. Since in this case the density of produced partons is low, the typical (semi-) hard partons produced in the initial scattering escape the interaction region without final state effects significantly affecting their back-to-back correlation. Considering on the other hand soft particle production amidst large parton densities in nucleus-nucleus collisions, it is well established that the azimuthal anisotropy of say $p_T \lesssim 1~\rm{GeV}$ particles is dominated by the final state response to the initial state geometry. In this case the mean-free path of a typical (semi-) hard parton is small compared to the system size, such that the initial state momentum correlations of $\sim\rm{GeV}$ partons are destroyed during the equilibration process. Therefore, the subsequent dynamics of the equilibrated QGP can be accurately described by relativistic hydrodynamics. 

Even though it is sometimes possible to choose the kinematics such that one mechanism dominates over the other, there are various examples in-between where both initial state and final state effects are important.  One prominent example includes the behavior of jets in heavy-ion collisions. While highly energetic jets can escape the interaction region without equilibrating, they can loose a significant part of their energy through interactions with the softer medium. Even though the dominant correlation of the leading high-$p_T$ particles is still due to the initial back-to-back correlation, the path length dependence of the energy loss in the medium also leads to an additional correlation with the initial state geometry. Such correlations are reflected e.g. by the high-momentum $v_{n}(p_T)$ measuring correlations between soft and hard particles.

\begin{figure}[t]
\includegraphics[width=0.5\textwidth]{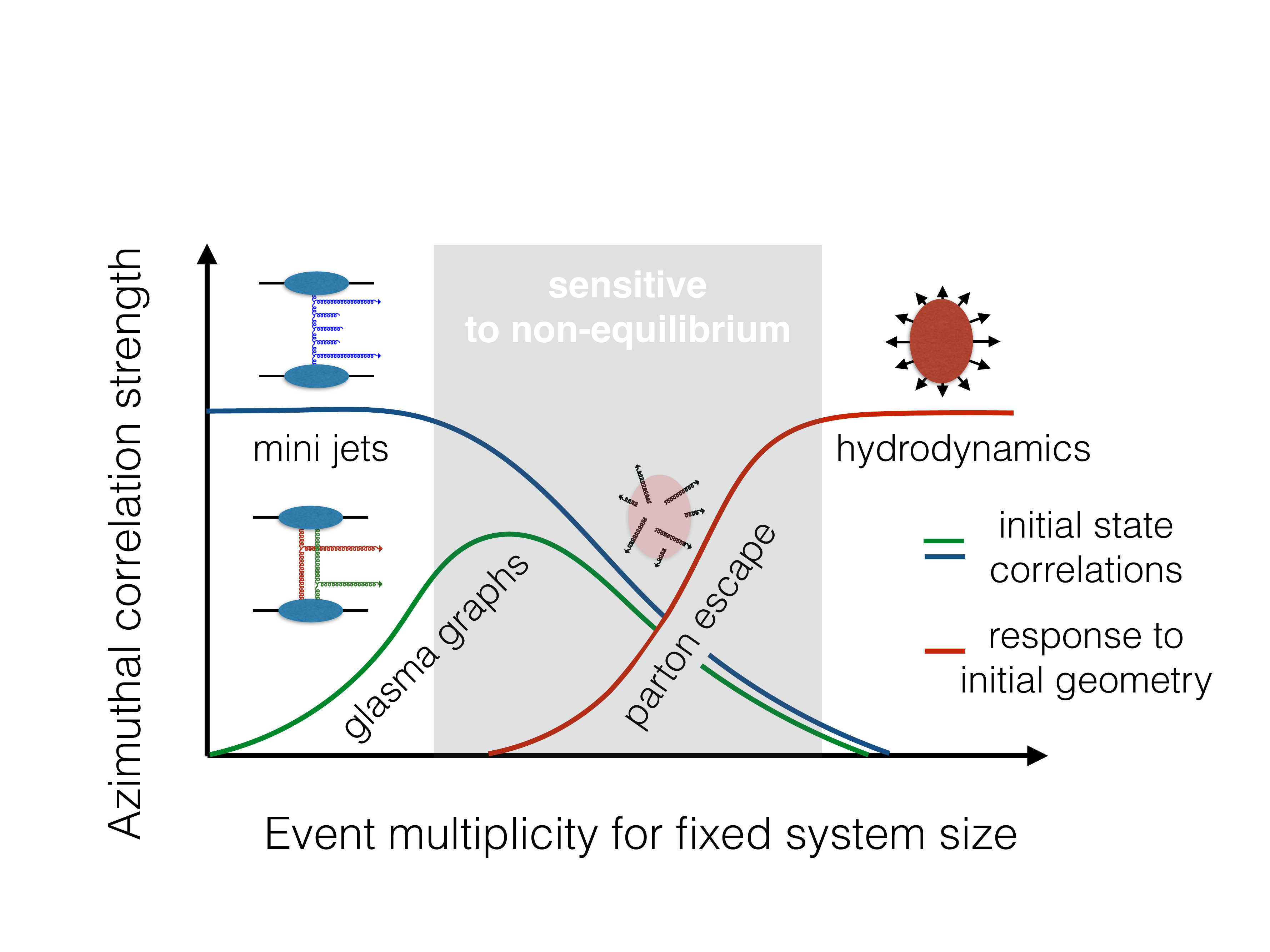} 
\caption{\label{PhaseDiagram} Illustration of long-range azimuthal correlations in small systems, a slightly modified version of the figure from~\cite{Schlichting:2016xmj}.}
\end{figure}

Clearly the aforementioned examples illustrate that it is important to consider both initial state momentum space correlations and the response to the initial state geometry due to final state effects in order to describe azimuthal correlations in small systems over a wide kinematic range. Our qualitative expectation is illustrated in Fig.~\ref{PhaseDiagram}, where the azimuthal correlation strength due to initial state and final state effects is shown versus the event multiplicity e.g. in $p+p$ collisions for a fixed transverse momentum range e.g. $1\!-\!3~\rm{GeV}$. Based on our discussion we expect that in low multiplicity or min-bias events the azimuthal correlations between $1\!-\!3~\rm{GeV}$ particles are pre-dominantly due to back-to-back mini-jets (peaked at $\Delta\phi=\pi$). With increasing event-multiplicity the contribution from multi-parton processes, such as the "Glasma graphs" (Sec.~\ref{sec_glasma}), becomes increasingly important resulting in azimuthal correlations that have a symmetric structure in relative azimuthal angle $\Delta\phi$ around $\pi/2$. When increasing the multiplicities even further, final state interactions in this transverse momentum region can no longer be neglected at some point  and lead to a depletion of initial state correlations. Even though mini-jets do not fully equilibrate yet, the system starts to show a response to the initial state geometry, which in this low opacity region is presumably dominated by the path length dependence of the parton energy loss -- also referred to as parton escape mechanism \cite{He:2015hfa}. Ultimately, in the limit of very high multiplicities, mini-jets are fully quenched, resulting in the formation of a thermalized medium and the complete loss of initial state momentum space correlations. In this high opacity regime, azimuthal correlations are dominated by the response to initial geometry described by a hydrodynamic expansion of a thermalized Quark-Gluon plasma.

One can attempt to further estimate the multiplicities corresponding to the transitions from the initial state to the final state dominated regime, exploiting recent theoretical progress in the understanding of the equilibration process~\cite{teaney2016is}. Since the equilibration time at weak coupling corresponds to the time scale when a semi-hard parton $\sim Q_s$ looses all its energy to form a soft thermal bath, one naturally expects the cross-over from the initial state to final state dominated regime to occur when the associated equilibration time $\tau_{eq}$ becomes comparable to the system size $R$. Conversely, as long as $\tau_{eq} \gg R$ typical semi-hard partons escape without encountering significant final state interactions, whereas for $\tau_{eq} \ll R$ semi-hard partons are fully quenched, equilibrium is reached early on and the dynamics is dominated by the subsequent hydrodynamic expansion. Based on the estimate of the equilibration time $Q_{s} \tau_{eq} \simeq 10 (\eta/s)_{T_{eq}}^{4/3} (g^2 N_c)^{1/3} \simeq 10 $ for  $(\eta/s)_{T_{eq}}\simeq 5/4\pi$ at realistic coupling $g^2N_{c}\simeq10$ ~\cite{Kurkela:2015qoa, Keegan:2016cpi} and the multiplicity $dN/dy\simeq \xi Q_s^2 \pi R^2$ with $\xi \simeq 1/4$ ~\cite{Lappi:2011gu} we obtain that 
\begin{eqnarray}
\label{eq:Crossover}
\frac{\tau_{eq}}{R} \simeq \sqrt{ \frac{100}{dN/dy}}\;,
\end{eqnarray}
corresponding to a cross-over at around $dN/dy\sim 100$, which in fact is much larger than the min-bias multiplicities reached in $p+p$ or $p+Pb$ collisions~\footnote{The typical values of $dN/dy$ in min-bias p+p and p+Pb collisions at LHC are about $\sim$ 6 and $\sim$ 17 ~\cite{Adam:2015gka, ALICE:2012xs} respectively.}. We caution however that the estimate in Eq.~(\ref{eq:Crossover}) is inferred from leading order weak-coupling calculations and should only serve as a ballpark figure. 

Beyond simple analytic estimates probably a promising alternative approach is to directly attempt an extraction of the boundaries between the different regimes through detailed comparisons of theory and experiment. While a first principle theoretical description is complicated throughout most of the multiplicity regimes shown in Fig.~\ref{PhaseDiagram}, significant theoretical progress has been made in understanding the features of initial state correlations in the regime where final state effects can be neglected. In the following we will review the theoretical computation of initial state correlations in the Color-Glass Condensate (CGC) effective field theory of high-energy QCD and critically access to what extent these calculations are compatible with the experimental observations.
\section{Multi-particle production in the CGC framework}

\begin{figure*}[t]
\hspace{-10pt}
\includegraphics[trim={0.1cm  0 0.2cm 0}, clip, width=0.3\textwidth]{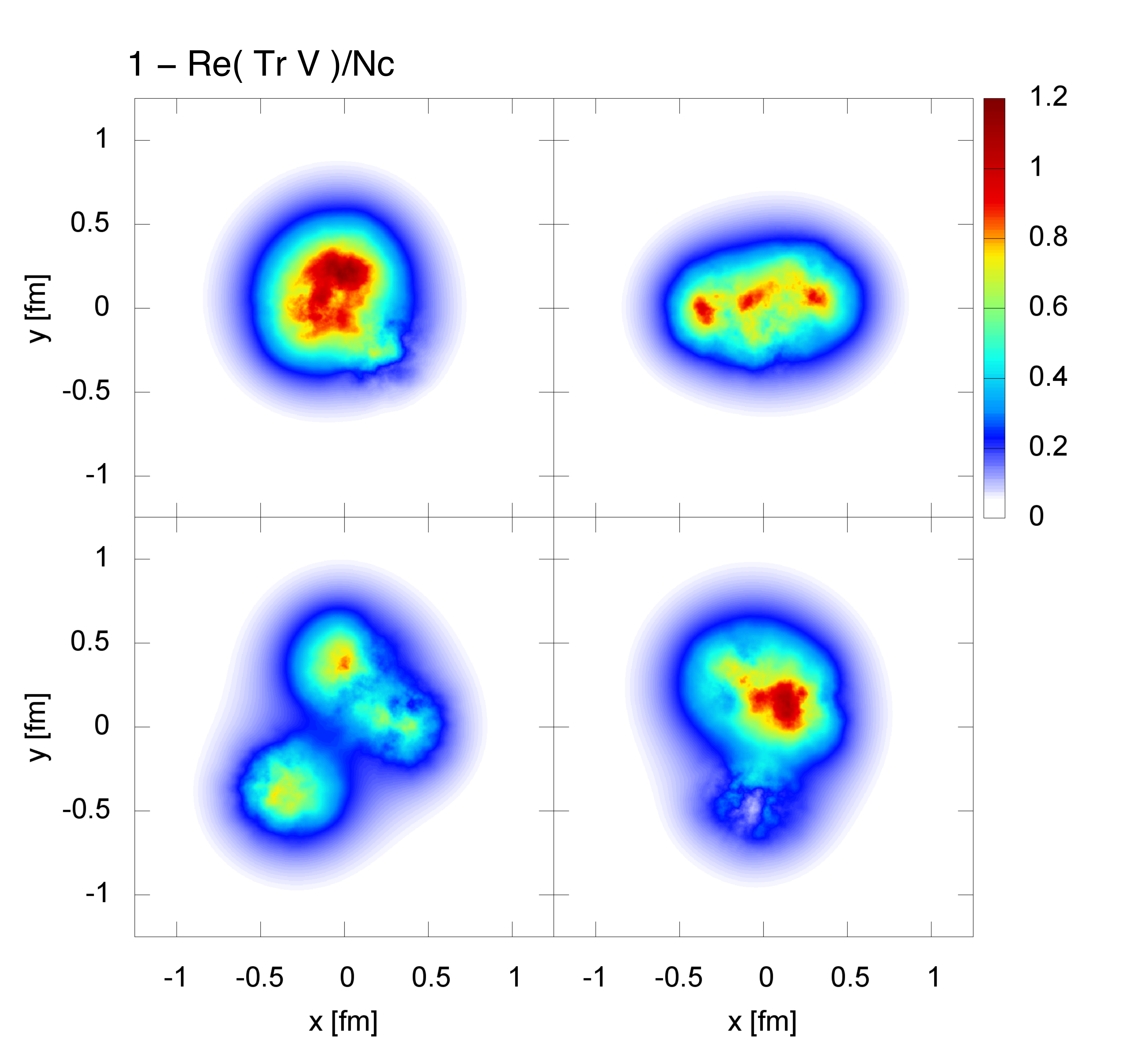} 
\includegraphics[trim={0.45cm 0 0 0}, clip, width=0.35\textwidth]{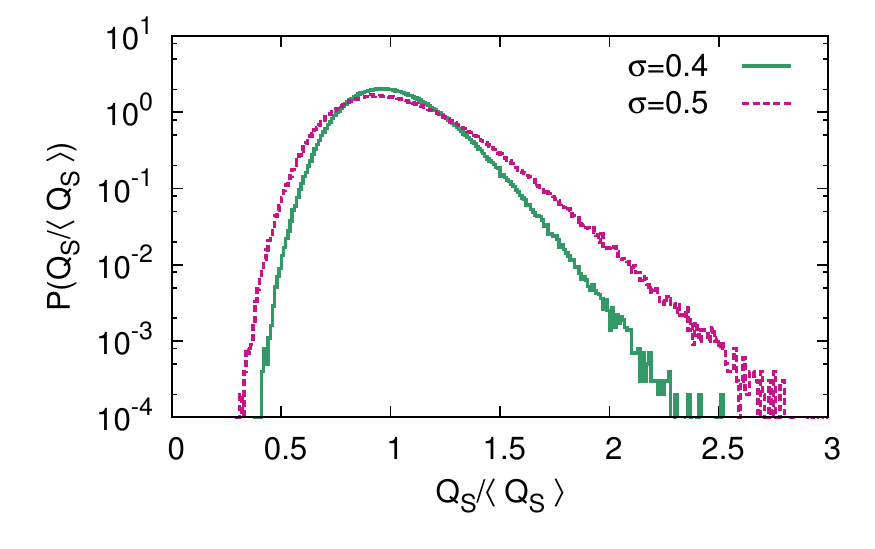}
\includegraphics[trim={0 0 0 0}, clip, width=0.35\textwidth]{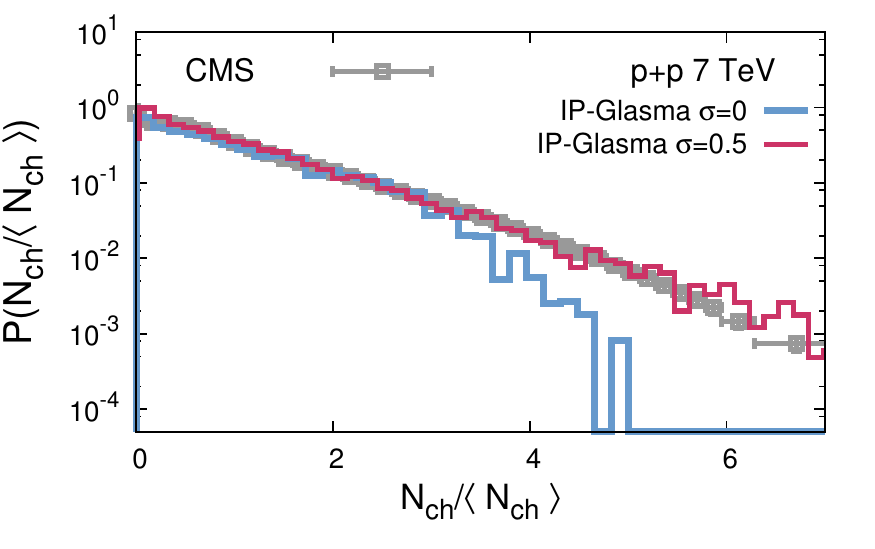}
\caption{\label{mult_fluc} (left) Distribution of gluon density around valance quarks inside a proton represented by the real part of Wilson lines from ~\cite{Mantysaari:2016ykx}.  (middle) Distribution of the saturation scale of a proton driven by stochastic dipole splitting. (right) Effect of intrinsic fluctuation of proton saturation scale on the probability distribution of multiplicity ($\sigma=0$ corresponds to no fluctuation of $Q_s$). The middle and the right figures are from Ref \cite{McLerran:2015qxa}.}
\end{figure*}

\subsection{High multiplicity events}
Experimental observations suggest that long-range ridge like correlations in small colliding systems appear in high multiplicity events. Before we turn to a more detailed discussion of possible mechanisms to produce such correlations, a first necessary step is to understand the origin of high multiplicity events that populate the long tail of experimental multiplicity distributions. Considering the most elementary case of p+p collisions, high multiplicity events are a consequence of three major sources of fluctuations
\begin{itemize}
\item[1)] geometry of collisions
\item[2)] intrinsic saturation scale of the proton
\item[3)] distribution of color charge density inside the protons.
\end{itemize}
While one naturally expects a strong impact parameter dependence of the multiplicity in $p+p$ collisions~\cite{d'Enterria:2010hd}, the importance of additional sources of initial state fluctuations have only been realized recently. Significant progress in including all of the above into a consistent phenomenological description has been made within the IP-Glasma model which is based on the framework of CGC~\cite{Schenke:2012wb,Schenke:2012hg}.

Non-perturbative large $x$ effects, which are not captured in the conventional CGC framework, are expected to give rise to fluctuations of the intrinsic saturation scale of the proton~\cite{Iancu:2004es, Iancu:2004iy, Munier:2003sj, Munier:2003vc, Mueller:2014fba, Marquet:2006xm} as illustrated in Fig.\ref{mult_fluc}\,(middle). In \cite{McLerran:2015qxa} intrinsic fluctuations of the proton saturation scale were introduced in the IP-Glasma model~\cite{McLerran:2015qxa} according to a distribution 
\beq
P(\ln(Q_s^2/\langle Q_s^2 \rangle)) =\frac{1}{\sqrt{2\pi}\sigma}\exp\left(  -\frac{\ln^{2}(Q_{s}^2/\langle Q_{s}^{2} \rangle)}{2\sigma^{2}}\right).%
\label{eq_qsdist}
\eeq
with the variance $\sigma^2$ adjusted previously to the inclusive charged particle multiplicity and rapidity distributions in p+p collisions over a wide range of energies 0.2-7 TeV~\cite{McLerran:2015qxa}.  An interesting consequence of such fluctuations is that even in symmetric collision systems such as $p+p$, event-by-event fluctuations of the saturation scale of each proton lead to an asymmetry of the rapidity distribution of the produced particles on an event-by-event basis~\cite{Bzdak:2015eii}. Consequently, high-multiplicity p+p collisions in particular are always asymmetric, and in a sense expected to look more like p+A collisions. 

The saturation scale and the geometric profile of the proton, which are the two most important ingredients in the IP-Glasma model are obtained from the IP-Sat parameterization of the HERA DIS data~\cite{Kowalski:2006hc,Rezaeian:2012ji}. While in the original IP-Sat model the proton shape was assumed to be round, recent modification to IP-Sat has been done by assuming the gluon density to be distributed around three valence quarks inside the proton (see Fig.\ref{mult_fluc} (left) )~\cite{Schlichting:2014ipa, Mantysaari:2016ykx} and several other models have started to include similar kinds of fluctuations~\cite{Welsh:2016siu, Bozek:2016kpf}. Interestingly such sub-nucleon scale fluctuations can be constrained by incoherent diffractive vector meson production and improved agreement with existing HERA data can be achieved~\cite{Mantysaari:2016ykx}. It has also been pointed out that geometric fluctuations of the proton can provide a dynamical explanation for the ``hollowness effect" observed in elastic p+p scattering at LHC energies~\cite{Albacete:2016pmp} and it would further be interesting to explore to what extent the modeling of the proton geometry can be aided by first-principle lattice QCD calculations.

\begin{figure*}[t]
\includegraphics[width=0.8\textwidth]{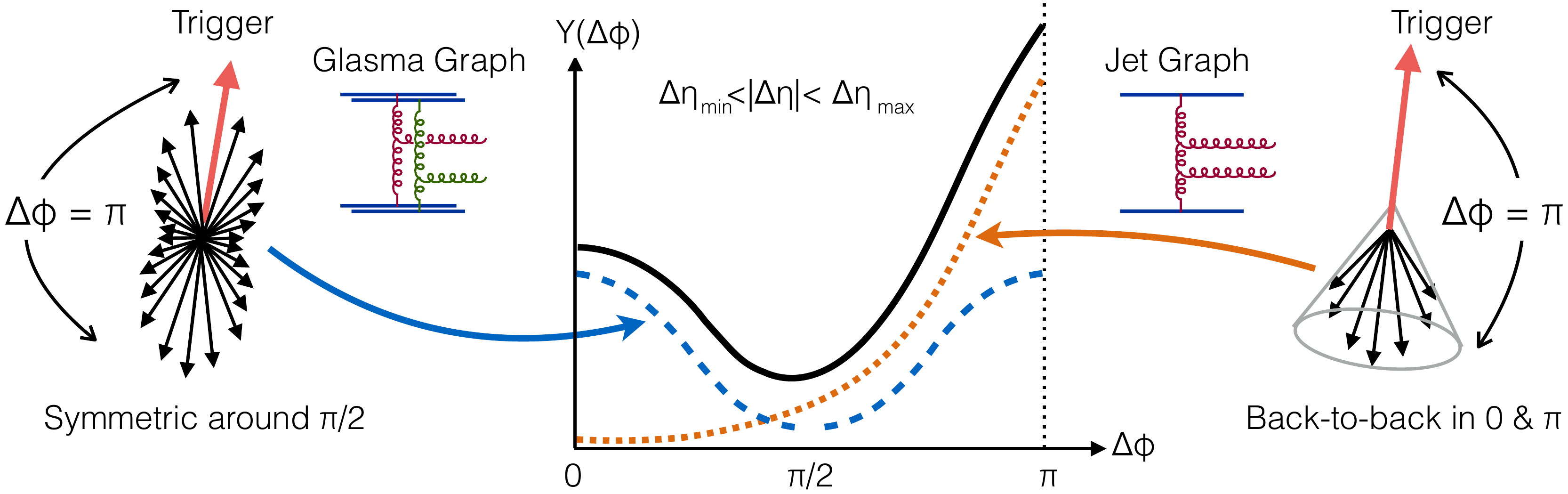}
\caption{\label{fig_ridge}A cartoon showing the contributions of di-jet and glasma graphs in two particle correlation function $Y(\Delta\phi)$ integrated over a broad range of $|\Delta\eta|$. This is a slightly modified version of the figure from ~\cite{Tribedy:2016tos}}
\end{figure*}

With the initial state including different sources of fluctuations  constrained by the HERA data, the $n$-particle production probability $P(n)$ can be computed within the color glass condensate framework. For a given configuration of initial color charge, the n-particle distribution is a negative binomial distribution (NBD) with mean and width related to the saturation scale~\cite{Gelis:2009wh}. Due to fluctuation of impact parameter, a convolution of many such NBDs gives rise to the final probability distribution of multiplicity. However it has been demonstrated that such distribution is narrower compared to the data. Only after including the intrinsic fluctuations of the proton saturation scale one can describe the tails of the experimental multiplicity distributions (Fig.\ref{mult_fluc}(right)). Since multiplicity is dominated by low momentum ($p_T<1$GeV) gluons, it is very challenging to implement a scheme of fragmentation for the production of soft hadrons. Therefore the number of produced charged particles are generally taken to be proportional to the number of gluons. The results shown in Fig.\ref{mult_fluc} indicate that the high multiplicity events that populate the tail of $P(n)$ distributions are generated due to rare high color charge density configurations of the wave functions of the colliding systems. In the next section we argue that the same underlying dynamics that leads to the origin of high multiplicity events also drives the systematics of multi-particle correlations in small collision systems. 

\subsection{Qualitative discussion of Initial state correlations}
Multi-particle production in Quantum Chromo Dynamics (QCD) naturally leads to correlations between particles produced in high-energy collisions. A complete theoretical understanding of these effects though is extremely challenging. Nevertheless significant progress has been achieved in recent years based on the CGC effective field theory (EFT) of high-energy QCD, which provides the basis for phenomenological applications at RHIC and LHC energies. 

Let us focus our discussion on the origin and systematics of the two particle correlations seen at LHC. By far the most well established source of long-range two-particle azimuthal correlations is due to the production of back-to-back di-jets.  Such processes (also referred to as ``Mueller-Navelet" jets~\cite{Mueller:1986ey}) are depicted in the right panel of Fig.~\ref{fig_ridge} and can be computed within standard perturbative QCD. Di-jet production is kinematically constrained to produce only away side (peaked at $\Delta\phi=\pi$) collimations and dominates in low-multiplicity or min-bias events. However, in high-multiplicity events one is probing rare configurations of the proton where in addition to the production of di-jets from a single hard scattering, multi-parton processes become increasingly important. A first calculation of these effects in the CGC framework was based on evaluating the associated Feynman diagrams referred to as ``Glasma graphs", depicted in the left panel of Fig.~\ref{fig_ridge}. Such graphs give rise to non-factorizable two particle correlations that have a symmetric structure in relative azimuthal angle $\Delta\phi$ around $\pi/2$ (see Fig.\ref{fig_ridge}). When decomposed in terms of the Fourier coefficients of the particle distributions, they give rise to non-zero even harmonics $v_{n}$. Beyond the lowest order processes depicted in the left panel of Fig.~\ref{fig_ridge}, further contributions to the azimuthal collimations come from the multiple scattering of partons  leading to both even and odd $v_{n}$. Such processes can be included in a classical Yang-Mills description and will be discussed in more detail in a following section.

Since interference effects between Glasma graphs and Jet graphs vanish to lowest order in the kinematic regime $\Lambda_{\rm QCD} \ll Q_s \lesssim p_T,q_T$ the resulting two-particle correlations function as a direct sum of both contributions
\begin{equation}
\frac{d^2N^{\rm \sl corr.}}{d^2\pp d^2\qp dy_p dy_q} = \frac{d^2N_{\rm \sl Glasma}^{\rm \sl corr.}}{d^2\pp d^2\qp dy_p dy_q} + \frac{d^2N_{\rm \sl Jet}^{\rm \sl corr.}}{d^2\pp d^2\qp dy_p dy_q}.
\end{equation}
The relative strength of the di-jet production represented by the ``Jet-graph" and the ``Glasma-graphs" determines the features of the observed di-hadron correlations as shown in Fig.\ref{fig_ridge}. In high multiplicity events the Glasma graphs are enhanced by a relative factor of $\alpha_S^{-4}$ compared to the ``Jet-graphs", one therefore naturally expects to see a pronounced near side collimation at $\Delta\phi\sim0$ that extends over a wide range of rapidity referred to as the ``near side ridge". The fact that near side collimation extends far in rapidity is a consequence of the nearly boost invariant nature of the glasma gluon fields. While qualitatively these features are indeed present in the experimental data, of course it requires detailed theory calculations to establish the quality of agreement. In the remainder of this section, we will outline the essential steps in the computation of initial state correlations in the CGC framework. A summary of comparisons with experimental results is presented in section~\ref{seq::comparison}.

 In the CGC framework colliding protons and nuclei are effectively described as static sources of color charge on the light-cone that generate color currents
\beq
J^\nu = \delta^{\nu\pm} \rho_{A(B)} (x^{\mp}, \xt).
\eeq
The color charge densities $\rho_{A(B)} (x^{\pm}, \xt)$ in each colliding hadron or nucleus fluctuate from event to event and their statistical properties are constrained by independent measurements. Computation of multi-particle production in the CGC framework is based on the calculation of the classical Yang-Mills fields created from such color currents by solving the Yang-Mills equations 
\beq
[D_\mu, F^{\mu\nu}] = J^\nu.
\label{eq_cym}
\eeq
Different theoretical descriptions within the CGC framework, employ different levels of approximation which are discussed in more detail in the following.

\subsection{Perturbative computation}
\label{seq_pertub}

In the perturbative framework one tries to obtain an analytical solution of the gauge fields by performing order-by-order expansion of Eq.\ref{eq_cym} in powers of the color sources $\rho_{A(B)}$. In the dilute-dense framework which assumes lowest order in $\rho_{A}$ and all orders in $\rho_{B}$ or in the dilute-dilute framework one can derive analytical expressions for n-gluon production in the $k_\perp$-factorized form~\cite{Blaizot:2004wu}. The essential ingredients to such factorization relations are the correlator of the dilute-sources in $\rho_{A}$ or the Wilson lines corresponding to the dense source $\rho_{B}$ in momentum space. Such correlators are represented as unintegrated gluon distributions (UGDs) $\Phi (x,k_\perp)$ and can be expressed in terms of ${\cal T}(x,r_\perp)$--the forward scattering amplitude of a quark-antiquark dipole of transverse size $r_\perp$ on a proton/nuclear target-- through the expression
\begin{equation}
\Phi(x, k_\perp) = {\pi \nc k_\perp^2\over 2\as}\int \limits_0^\infty dr_\perp r_\perp J_0(k_\perp r_\perp)  [1-{\cal T} (x,r_\perp)]^2.
\label{eq_ugd}
\end{equation}

The $x$-dependence of $\Phi$ is determined by the rapidity ($Y=\ln(1/x)$) evolution of the dipole scattering amplitude  implemented in the Balitsky-Kovchegov (BK) renormalization equation~\cite{Balitsky:1995ub, Kovchegov:1999yj} which is a simplified form of the JIMWLK renormalization equations~\cite{Balitsky:1995ub, PhysRevD.61.074018, Jalilian-Marian:1997xn, Iancu:2000hn, Ferreiro:2001qy}. The leading order expression of the BK equation is given by 
\begin{eqnarray}
\frac{\partial T(\rb{},Y)}{\partial Y} 
=\int_{\rb{1}} {\mathcal K}(\rb{},\rb{1})
\left[ T(\rb{1},Y) + T(\rb{2},Y) \right. \\ \nonumber
\left.- T(\rb{},Y) - T(\rb{1},Y)\,T(\rb{2},Y)\right],
\label{eq_loBK}
\end{eqnarray}
where $\rb{2}\equiv \rb{}-\rb{1}$ and ${\mathcal K}$ is the BFKL kernel. The implementation of the kernel ${\mathcal K}$ often used for phenomenology includes a running coupling next-to-leading-log (NLL) correction to BK and is referred to as the rcBK equation~\cite{Balitsky:2006wa} given by 
\begin{eqnarray}
{\mathcal K}(\rb,\rb{1},\rb{2}) = \frac{\alpha_s(\rb{}) N_c}{\pi}\left[ \frac{\rb{}^2}{\rb{1}^2 \rb{2}^2} 
+\frac{1}{\rb{1}^2}\left(\frac{\alpha_s(\rb{1}^2)}{\alpha_s(\rb{2}^2)}-1\right) \right. \\ \nonumber
\left.+\frac{1}{\rb{2}^2}\left(\frac{\alpha_s(\rb{2}^2)}{\alpha_s(\rb{1}^2)}-1\right)\right] \,.
\label{eq_rcBK}
\end{eqnarray}
Eq.\ref{eq_loBK} requires an initial condition for ${\cal T}_{A,B} (x=x_0,r_\perp)$, one choice of which is 
the McLerran-Venugopalan (MV) model~\cite{McLerran:1993ni,McLerran:1993ka} with a finite anomalous dimension $\gamma$ given by 
\begin{equation}
{\cal T}(x_0, r_\perp) = 1-\exp\left[-\left(\frac{r_\perp^2Q_{s0}^2}{4}\right)^\gamma\ln\left(\frac{1}{r_\perp \Lambda_{QCD}}+e\right)\right].
\label{eq_mvic}
\end{equation}
where $Q_{s0}^2$ is a non-perturbative scale. This MV-like parameterization is constrained by global fits to the DIS data~\cite{Albacete:2010sy} although it must be noted that MV along with BK evolution does not include spatial geometric structure of the proton. The scattering amplitude is known to have a strong dependence on the impact parameter~\cite{Kowalski:2006hc} which is incorporated in other parameterization such as IP-Sat~\cite{Kowalski:2003hm, Rezaeian:2012ji} or b-CGC ~\cite{Kowalski:2006hc} models of DIS.  More recently, significant progress in consistently including the full NLL corrections to BK evolution into DIS fits \cite{Iancu:2015vea,Iancu:2015joa,Lappi:2016fmu} has been made. However, so far this has not been included in phenomenological studies of p+p and p+A collisions.

With the un-integrated gluon distribution obtained from Eq.\ref{eq_ugd}, one can estimate the production of n-gluons using the $k_\perp$ factorization approach. In the following section we describe the approach for single-inclusive, double-inclsuive and jet production essential for the phenomenology in p+p and p+A collisions.

\subsubsection{Single inclusive gluon distribution} 
The single inclusive gluon production corresponds to the simplest process that describes the emission of a single gluon of momentum $\pp$ which in the perturbative framework can be written in the $k_\perp-$ factorized form as 
\begin{align}
\frac{\ud N}{\ud y_p\ud^2\pp }
&=\frac{8\alpha_s }{(2\pi)^6 \cf}\frac{S_\perp}{\pp^2} 
\times \int\limits_{\kp} \Phi_{A}(\kp)\,\Phi_{B}(\pp-\kp)\,.
\label{eq:single}
\end{align}
where $\cf=(\nc^2-1)/2 \nc$ and $S_\perp$ is the transverse overlap area. It must be noted that this $k_\perp$-factorized form of single gluon  production has a logarithmic infrared divergence which is generally regulated by putting a lower $p_{T,\rm min}$ cut. It must be noted that Eq.\ref{eq:single} assumes that the dependence on transverse geometry or the impact parameter of collisions has already been integrated out and absorbed into $S_\perp$, a more general expression in such a context can be found in Ref~\cite{Blaizot:2004wu}. 

\subsubsection{Di-Jets}
In the perturbative framework, estimation of the two-particle correlations in $p+p$ and $p+A$ collisions are based on a direct computation of di-jet and the glasma graphs in the $k_\perp$-factorization approximation~\cite{Dusling:2012iga,Dusling:2012cg,Dusling:2012wy,Dusling:2013qoz,Dusling:2015rja}. Such approximations are valid for momenta above the saturation scale $Q_s$ and do not include multiple-scattering effects. 

\begin{figure}[t]
\includegraphics[width=0.5\textwidth]{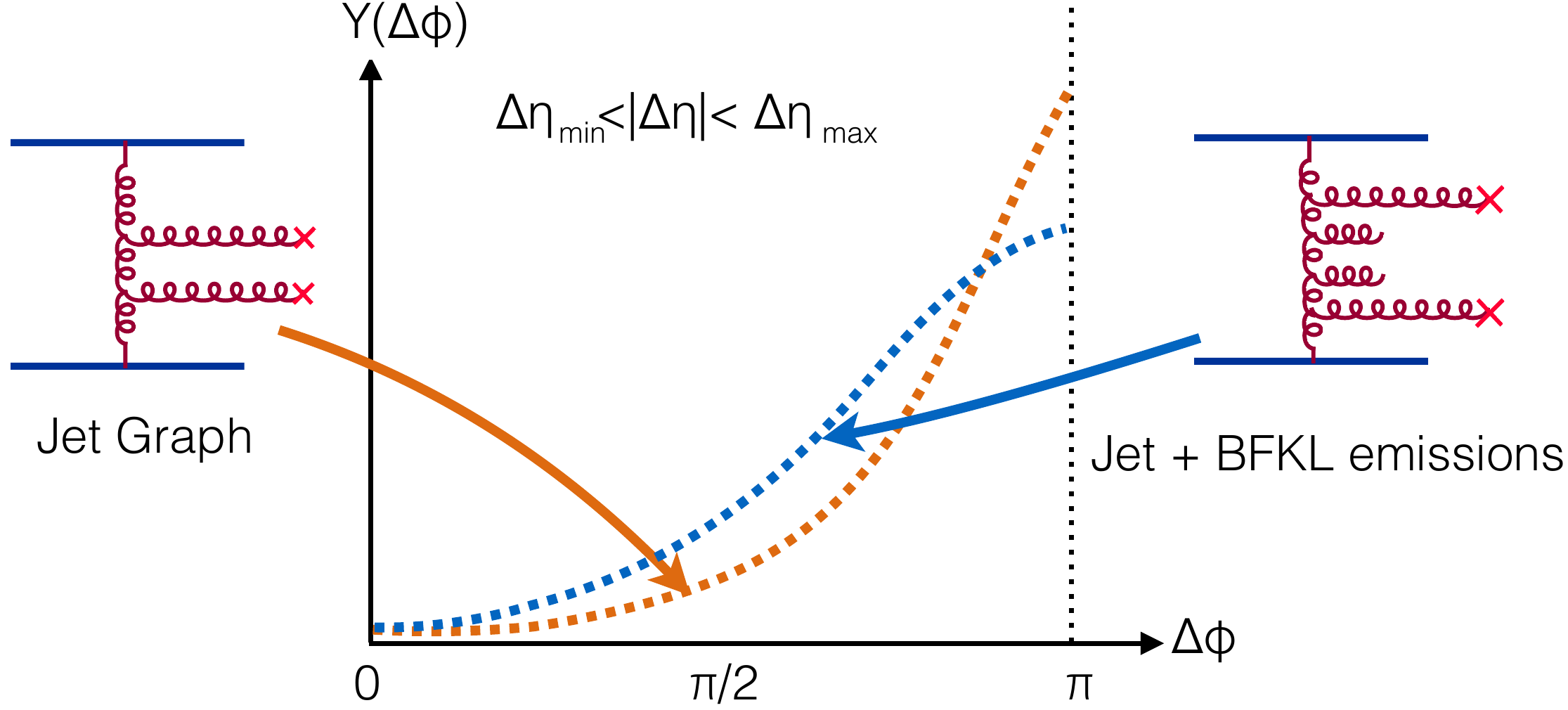}
\caption{\label{fig_bfkl}A cartoon showing the contributions of di-jet and glasma graphs in two particle correlation function $Y(\Delta\phi)$ integrated over a broad range of $|\Delta\eta|$.}
\end{figure}

The di-jet contribution in this framework is estimated ~\cite{Mueller:1986ey, Colferai:2010wu, Fadin:1996zv} to be 
\begin{eqnarray}
\label{eq:bfkl}
&&\frac{d^2N_{\rm \sl BFKL}^{\rm \sl corr.}}{d^2\pp d^2\qp dy_p dy_q} = \frac{32\,\nc\,
\alpha_s^2}{ (2\pi)^8 \,\cf}\,\frac{\sp}{\pp^2\qp^2}\times \\
&&\int \limits_{{\kpn{0}}} \! \int \limits_{{\kpn{3}}} \!\!
\Phi_A(\kpn{0})\Phi_B(\kpn{3})\,\mathcal{G}(\kpn{0}\!\!-\!\!\pp,\kpn{3}\!\!+\!\!\qp,y_p\!\!-\!\!y_q) ,\nonumber
\end{eqnarray}
where $\mathcal{G}$ is the BFKL Green's function that generates gluon emissions between the gluons that fragment into triggered hadrons. The form of $\mathcal{G}$ is given by ~\cite{Dusling:2013qoz}. 
\begin{eqnarray}
\mathcal{G}(\qpn{a},\qpn{b},\Delta y)= \frac{1}{(2\pi)^2}\frac{1}{(\qpn{a}^2 \qpn{b}^2)^{1/2}} \times \\ \nonumber
\sum_n e^{in\overline{\phi}}\int_{-\infty}^{+\infty} d\nu\textrm{ } e^{\omega(\nu,n)\Delta y}e^{i\nu\ln\left(\qpn{a}^2/\qpn{b}^2\right)}\textrm{   } \, .
\label{eq:BFKL-Green}
\end{eqnarray} 
Here ${\omega(\nu,n)=-2\overline{\alpha}_s\,
\textrm{Re}\left[\Psi\left(\frac{|n|+1}{2}+i\nu\right)-\Psi(1)\right]}$ is the
BFKL eigenvalue with $\Psi(z)= d\ln\Gamma(z)/dz$ being the logarithmic
derivative of the Gamma function with the effective coupling factor $\overline{\alpha}_s\equiv
\nc\,\as\left(\sqrt{\qpn{a}\qpn{b}}\right)/\pi$ and 
$\overline{\phi}\equiv \arccos\left(\frac{\qpn{a}\cdot \qpn{b}}{\vert\qpn{a}\vert\textrm{ }\vert \qpn{b}\vert}\right)$. 
For a pair of hadrons with a rapidity separation of $\Delta y\gtrsim 1/\alpha_s$,  $\mathcal{G}$ does the necessary resummation of the rapidity ordered multi-gluon emissions. The effect of such gluon emission leads to angular de-correlation that affects the observed di-hadron correlation~\cite{Mueller:1986ey}. The diagrams corresponding to such BFKL emission and its impact on broadening the away-side structure of the azimuthal di-hadron correlation is shown in Fig.\ref{fig_bfkl}.  
In the $\as\Delta y\to 0$ limit one can obtain the well known form of the di-jet cross-section expression in the Multi-Regge kinematics (MRK)~\cite{Fadin:1996zv,Leonidov:1999nc} : 
\begin{eqnarray}
\left.\frac{d^2N_{AB}}{d^2\pp d^2\qp dy_p dy_q}\right\vert_{\rm MRK} = 
\frac{16\,\nc\, \alpha_s(\pp)\,\alpha_s(\qp)}{ (2\pi)^8 \,\cf}\, \\ \nonumber
\times \, \frac{\sp}{\pp^2\qp^2}\int_{\kpn{1}} \Phi_A(x_1,\kpn{1})\Phi_B(x_2,\kpn{2})\,.
\label{eq:MRK}
\end{eqnarray}
As we discuss in the following section, the quantitative estimation of the di-jet cross-section is essential for the description of the di-hadron correlation observed in high multiplicity events of p+p and p+Pb collisions. 

\subsubsection{Glasma graph}
\label{sec_glasma}

In the context of two particle correlations, there are a total of eight topologies of the glasma graph (full expression can be found in Ref~\cite{Dusling:2013qoz}), the contribution for one such diagram to the two-particle correlations can be written as 
\begin{align}
\label{eq:glasma}
\frac{d^2N_{\rm \sl Glasma-1}^{\rm \sl corr.}}{d^2\pp d^2\qp dy_p dy_q}&=
\frac{32\as^2}{(2\pi)^{10}\zeta\;\nc\cf^3}\,\frac{S_\perp}{\pp^2\qp^2}\times\\
& \int \limits_{\kp} \Phi_A^2(\kp)\Phi_B(\pp-\kp)\Phi_B(\qp-\kp)\;.\nonumber 
\end{align}
Beyond the straightforward, perturbative result the non-perturbative factor $\zeta$ has been introduced to account for the contributions to multi-particle production below the scale $\qs^2$. Constrains from the independent analysis of $n$-particle multiplicity distributions~\cite{ Lappi:2009xa, Tribedy:2011aa, Tribedy:2010ab, Schenke:2012fw} suggest that $\zeta$ be in the range $0.1\!-\!1$ and phenomenological studies typically employ the value of $\zeta=1/6$ obtained from ~\cite{Tribedy:2011aa, Tribedy:2010ab}.  

%

The collimation in the glasma graph framework comes from the fact that each of the four UGDs in Eq.\ref{eq:glasma} has a bell-shaped structure with a maximum corresponding to the saturation momentum. Such nature of the UGDs kinematically constrains two gluons to be produced in similar (or back-to-back) directions giving rise to the double ridge structure in two particle correlations~\cite{Dusling:2012iga, Altinoluk:2015uaa}. 

\subsection{Dilute-dense models}
Besides the perturbative diagrammatic approach taken in the Glasma graph calculation, several studies have been launched to investigate the origin of structure of the long-range azimuthal correlations in the limit where a dilute projectile of individual partons scatter off the color-fields of a dense projectile. Neglecting correlations of the incoming partons in the projectile, the double inclusive distribution of the scattered partons takes the form~\cite{Lappi:2015vha,Lappi:2015vta}
\begin{eqnarray}
\label{eq:dildenDoubleInclusive}
&&\frac{\ud^2N}{\ud y_1 \ud^2\ptv_1  \ud y_2 \ud^2\ptv_2}=  \int \ud^2\btv_1 \ud^2\btv_2 \int \frac{\ud^2\ktv_1}{(2\pi)^2} \int \frac{\ud^2\ktv_2}{(2\pi)^2}   \int \ud^2\rtv_1 \ud^2\rtv_2 \nonumber \\
&& \quad e^{i(\ptv_1-\ktv_1) \cdot \rtv_1} e^{i(\ptv_2-\ktv_2)\cdot \rtv_2} W_{q/g,\ud y_1}(\btv_1,\ktv_1)  W_{q/g,\ud y_2}(\btv_2,\ktv_2)  \nonumber \\
&& \quad \bigg< \mathcal{D}\left(\btv_1+\frac{\rtv_1}{2},\btv_1-\frac{\rtv_1}{2}\right)\mathcal{D}\left(\btv_2+ \frac{\rtv_2}{2},\btv_2- \frac{\rtv_2}{2}\right) \bigg> .
\end{eqnarray}
where $W_{q/g,\ud y_1}(\btv_1,\ktv_1) $ denotes the Wigner function of quarks/gluons inside the projectile and $\mathcal{D} (\textbf{x},\textbf{y}) = \frac{1}{N_c}  \tr \left[ V(\textbf{x}) V^{\dagger}(\textbf{y}) \right]$ is the usual dipole-correlator of light-like Wilson lines in the fundamental/adjoint representation.  While the transverse momenta $\ktv_1$ and $\ktv_2$ of the two incoming quarks are uncorrelated, it is evident from Eq.~\ref{eq:dildenDoubleInclusive} that the correlations in the momentum transfers $\ptv_1-\ktv_1$ and $\ptv_2-\ktv_2$ give rise to azimuthal correlations of the scattered partons $\ptv_1$ and $\ptv_2$. We note that in contrast to the perturbative calculation, present calculations in the dilute-dense description usually do not take into account rapidity evolution between the produced particles and are therefore limited to the kinematic range where $y_1-y_2 \ll 1/\alpha_s$ where evolution effects can be neglected. While recent progress has been made in formulating evolution equations for multi-particle production in dilute-dense systems \cite{Iancu:2013uva}, these have not been employed for phenomenological studies so far.

\begin{figure}[t]
\includegraphics[width=0.2725\textwidth]{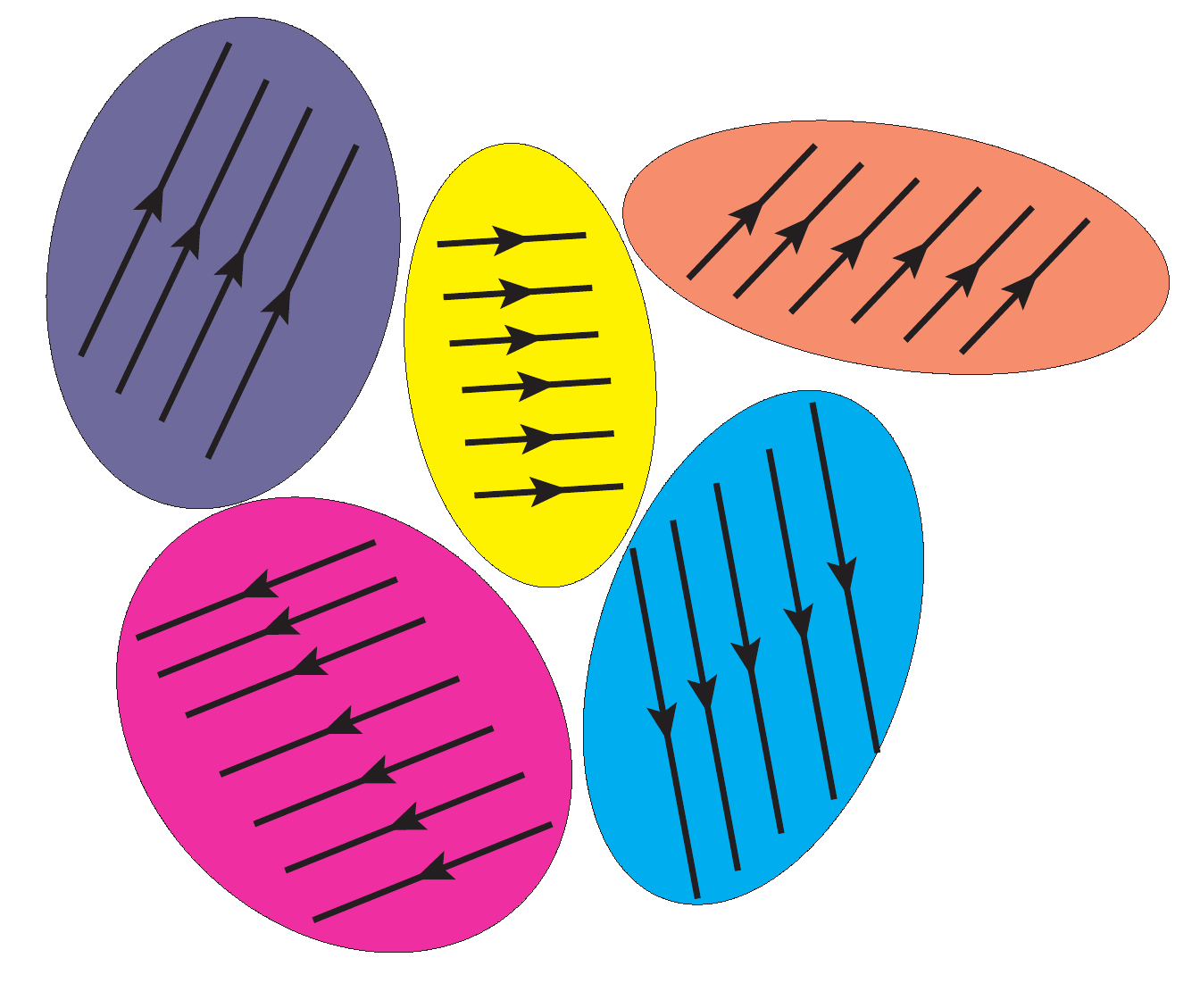}
\caption{\label{fig_domains} Color fields of the target are locally organized in domains of characteristic size of the inverse saturation scale $1/Q_s$,  taken from ~\cite{Lappi:2015vta}}
\end{figure}

By evaluating the dipole operator in short-distance expansion $(Q_s|\mathbf{x}-\mathbf{y}|<\!<1)$
\begin{eqnarray}
\label{eq:E-dipole}
\mathcal{D}(\textbf{x},\textbf{y}) \simeq 1 -  \frac{(x-y)^i (x-y)^j}{4\nc} E_i^a\Big(\frac{\textbf{x}+\textbf{y}}{2}\Big) E_j^a\Big(\frac{\textbf{x}+\textbf{y}}{2}\Big), \nonumber \\
\end{eqnarray}
where $E_i(\textbf{x}) = i V(\textbf{x})\partial_{i}V^{\dagger}(\textbf{x})$ denotes the light-cone electric field, one can establish an intuitive picture of the origin of the long-range near side correlations \cite{Kovner:2010xk}. When a projectile parton scatters off the color field of the nucleus it receives a transverse momentum kick in the direction of the color-electric field of the target. Color fields fluctuate from event to event and are locally organized in domains of size $\sim 1/Q_s$ as illustrated in Fig.~\ref{fig_domains}. When two (or more) quarks scatter off the same domain, they will receive a similar kick whenever they are in the same color state. Naturally, this leads to a correlation which is suppressed by $1/N_c^2$ (in the limit of large $N_c$) and by the number of domains $Q_s^2S_\bot$, where $S_{\bot}$ denotes the transverse area probed by the projectile.

Several studies have computed the azimuthal harmonics of the two-particle correlation function in Eq.~(\ref{eq:dildenDoubleInclusive}), either based directly on numerical evaluations including JIMWLK evolution \cite{Lappi:2015vha,Lappi:2015vta} or within (semi-)analytic models \cite{Dumitru:2014dra,Dumitru:2015cfa, McLerran:2015sva}. Generally the double-inclusive spectrum in Eq.~\ref{eq:dildenDoubleInclusive} features sizable azimuthal correlations $v_{n} $ which are on the order of $1/\sqrt{N_c^2-1}$ and most pronounced when both parton momenta $p_{1}$ are $p_2$ are on the order of the saturation scale. While the double-inclusive spectrum for the two incoming quarks features both even ($n=2,4,...$) and odd  ($n=1,3,5,...$) harmonics, it turns out that the odd harmonics for gluons vanish identically due an exact symmetry of the the correlation function under $\ptv_1 \to - \ptv_1$. Even though the dilute-dense framework described above certainly does not provide the most realistic initial state model for mid-rapidity particle production in high-multiplicity events, it turns out that the absence of odd harmonics for gluons poses a more general problem within the CGC framework. However, as discussed in Sec.~\ref{seq::comparison} recent simulations including the early-time classical Yang-Mills dynamics have been able to resolve this puzzle to some extent.

\subsection{Classical Yang-Mills}
Beyond the approaches outlined above there have also been new theoretical developments in the study of initial state correlations in event-by-event simulations in classical Yang-Mills theory \cite{Schenke:2012wb, Schenke:2012fw,Schenke:2015aqa}. In this approach  Eq.\ref{eq_cym} is solved numerically for the individual colliding hadrons or nuclei in Lorentz gauge $\partial_\mu A^\mu = 0$, where
\begin{equation}\label{eq:lor}
 A_{A(B)}^\pm = -\frac{\rho_{A (B)}(x^\mp,\xt)}{\boldsymbol{\nabla}_\perp^2+m^2}\,,
\end{equation}
and then transformed into the light-cone gauge $A^+ (A^-) = 0$, where one finds 
~\cite{McLerran:1994ni,*McLerran:1994ka,*McLerran:1994vd,JalilianMarian:1996xn,Kovchegov:1996ty}   
\begin{align}\label{eq:sol}
 A^i_{A (B)}(\xt) &= \theta(x^-(x^+))\frac{i}{g}V_{A (B)}(\xt)\partial_i V^\dag_{A (B)}(\xt)\,,\\
 A^- (A^+) &= 0\,.\label{eq:sol2}
\end{align}
The infrared regulator $m$ in Eq.\,(\ref{eq:lor}) is of order $\Lambda_{\rm QCD}$ and crudely incorporates color confinement at the nucleon level by damping the Coulomb tail. The gauge field in the forward light-cone after the collision at time $\tau=0$ is given by the solution of the CYM equations in Fock--Schwinger gauge $A^\tau=(x^+ A^- + x^- A^+)/\tau=0$ in terms of the gauge fields of the colliding nuclei~\cite{Kovner:1995ja,Kovner:1995ts}:
\begin{align}
 A^i &= A^i_{(A)} + A^i_{(B)}\,,\label{eq:init1}\\
 A^\eta &= \frac{ig}{2}\left[A^i_{(A)},A^i_{(B)}\right]\,,\label{eq:init2}\\
 \partial_\tau A^i &= 0\,,\\
 \partial_\tau A^\eta &= 0
\end{align}
Such gluon fields produced after the collision are evolved in time according to Classical Yang-Mills equations up to time $\tau\sim 1/Q_s$ to estimate the gluon spectrum $dN_g/dyd^{2}k_{t}$ by imposing Coulomb gauge $\left.\partial_{i}A^{i}\right|_{\tau}=0$ and extracting the equal time correlation function \cite{Berges:2013fga, Berges:2013eia}
\begin{eqnarray}\label{eq:singleParticleDist}
\left.\frac{dN}{d^{2}\kt dy}\right|_{\tau}=\frac{1}{(2\pi)^2} \sum_{\lambda,a} \left| \tau g^{\mu\nu} \Big( \xi_{\mu}^{\lambda,\kt*}(\tau) \overleftrightarrow{\partial_{\tau}} A_{\nu}^{a}(\tau,\kt)\Big) \right|^2 \nonumber \\
\end{eqnarray}
where  $g^{\mu\nu}=(1,-1,-1,-\tau^{-2})$ denotes the Bjorken metric and $\lambda=1,2$ labels the two transverse polarizations. In Coulomb gauge the mode functions take the form
\begin{eqnarray}
\xi_{\mu}^{(1),\kt}(\tau)&=& \frac{\sqrt{\pi}}{2|\kt|}  \begin{pmatrix} -k_y \\ k_x \\ 0 \end{pmatrix}  H^{(2)}_{0}(|\kt|\tau) \;, \label{eq:A1} \\
\xi_{\mu}^{(2),\kt}(\tau)&=&\frac{\sqrt{\pi}}{2|\kt|}  \begin{pmatrix} 0 \\ 0 \\  k_T\tau \end{pmatrix}  H^{(2)}_{1}(|\kt|\tau) \;, \label{eq:A2}
\end{eqnarray}
where $H^{(2)}_{\alpha}$ denote the Hankel functions of the second type and order $\alpha$ (see \cite{Berges:2013fga} for details).

The calculations in the framework of Classical Yang-Mills naturally include the Glasma graphs and extend the reach of perturbative calculations towards lower $p_T$ by consistently including multiple-scattering effects as well as coherent re-scattering in the final state. Since event-by-event simulations in the classical Yang-Mills theory also allow for an improved treatment of the impact parameter dependence, they can be used in the future to systematically study initial state effects across different collision geometries e.g. in $p+A$,~$d+A$ and $^3He+A$ collisions at RHIC \cite{Adare:2013piz,Adare:2015ctn,Adamczyk:2015xjc}.

\begin{figure*}[ht]
\includegraphics[width=0.65\textwidth]{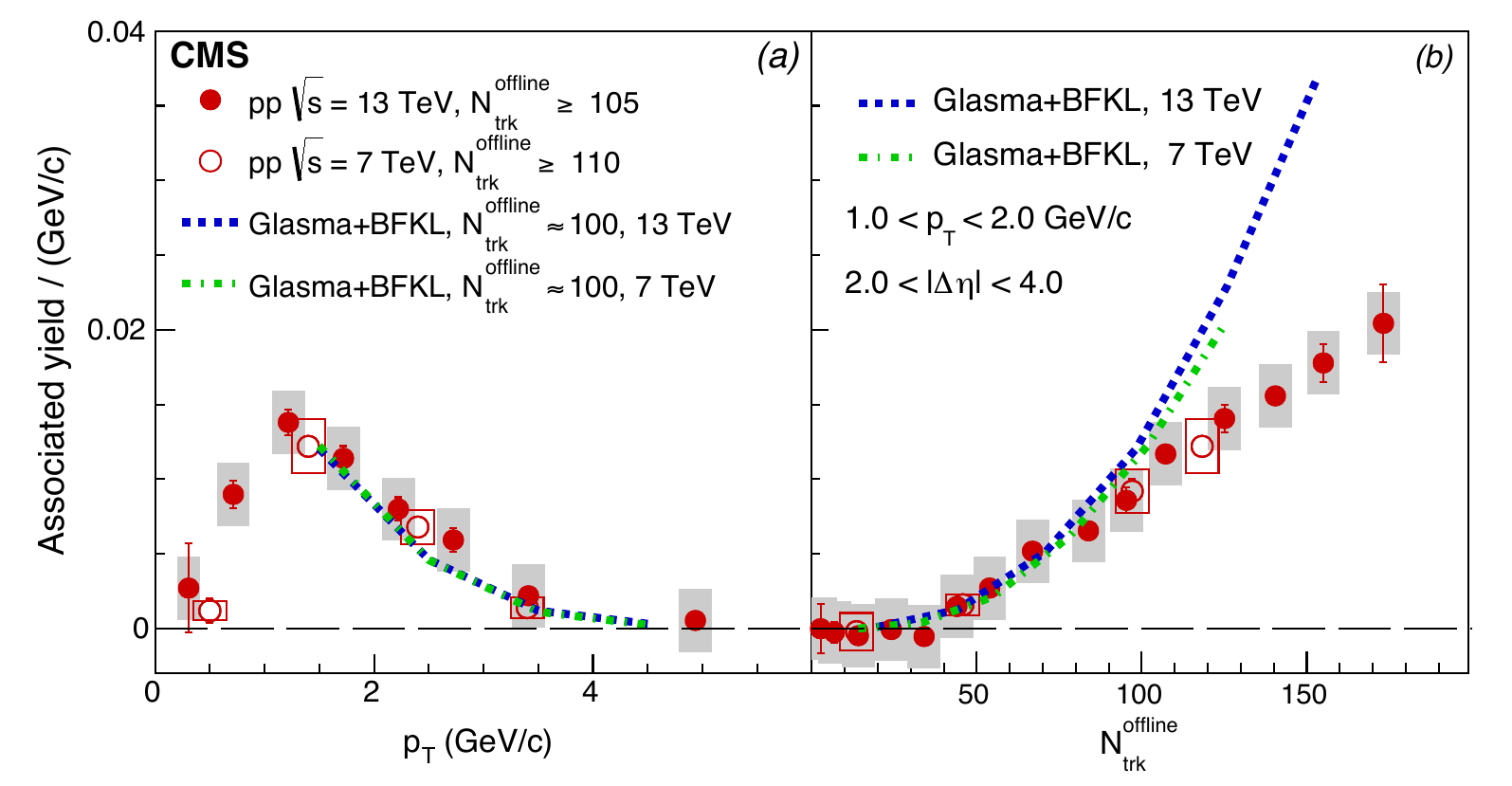}
\caption{\label{fig_scaling} Figure taken from Ref.~\cite{Khachatryan:2015lva} showing the measurements by the CMS collaboration on the transverse momentum dependence of the near side ridge yield (left) and the multiplicity dependence of the same quantity shown for two energies in p+p collisions (right). Model calculations (shown by dashed lines) are from Ref.~\cite{Dusling:2015rja} using the Glasma graph and BFKL approach discussed in section \ref{seq_pertub}.}
\end{figure*}

\subsection{Hadronization}
So far we have outlined the computation of initial state correlations at the parton level. The mechanism of hadronization converts the partonic correlations driven by initial state dynamics into correlated production of final-state particles. Implementation of a realistic scheme of hadronization is therefore essential for the phenomenology of small systems collisions. A first principle QCD based approach to such a problem is very challenging. One therefore resorts to several available approximation schemes for fragmentations of partons. The most commonly used approach is the standard  parton-hadron independent hadronization scheme~\cite{Field:1977fa} in which e.g. the single inclusive hadron distributions are obtained by convoluting the gluon distributions with fragmentation functions as  
\begin{equation}
\frac{d{N}_{h}}{d^2\pp dy} = \int_{z_{\rm min.}}^1 \frac{dz}{z^2}\, \frac{d{N}_{g}}{d^2\qp dy}\,D_{g\rightarrow h}\left(z=\frac{\pp}{\qp}, \mu^2\right) \,,
\label{eq:single-inclusive}
\end{equation}
where $D_{g\rightarrow h}(z,\mu^2)$ denotes the probability that a gluon fragments into a hadron carrying $z$ fraction of its momentum at a scale $\mu^2$. The lower limit of the integral is determined from the kinematic requirement that the momentum fraction of the gluons  $x\le1$. Commonly used form for $D_{g\rightarrow h}(z,\mu^2)$ such as KKP or DSS fragmentation functions are obtained from fits to the inclusive hadron production data in $e^+\!+\! e^-$ and p+p collisions~\cite{Kniehl:2000fe, deFlorian:2007aj}. In case of double inclusive production relevant for the study of two-particle correlations, one assumes an ansatz of the form~\cite{Dusling:2012iga}
\begin{align}
\label{eq:dihadron}
& \frac{d^2N_{h}^{\rm \sl corr.}}{d^2p_T d^2q_T d\eta_p d\eta_q}= \\
&\!\! \int\limits_{z_0}^1\!\! dz_1 dz_2 \frac{D(z_1)}{z_1^2}\, \frac{D(z_2)}{z_2^2}
 \frac{d^2N_{g}^{\rm \sl corr.}}{d^2\pp d^2\qp d\eta_p d\eta_q}\left(\frac{p_{\textrm{T}}}{z_1},\frac{q_{\textrm{T}}}{z_2} \right).\nonumber
\end{align}
A limitations of is standard approach of hadronization is that the applicability of the fragmentations functions are questionable at low virtuality $\mu^2$. Therefore such scheme of hadronization can not be used for bulk particle production which is dominated by soft processes with typical virtuality scale $\mu^2 \lesssim 1$ GeV. In this regime an alternative approach such as the Local Parton Hadron Duality (LHPD)~\cite{Azimov:1984np} can be used to describe bulk particle production driven by initial sate dynamics~\cite{Levin:2010dw}. 

The state-of-the-art scheme of hadronization to describe bulk particle production used in several event generators like PYTHIA~\cite{Andersson:1983ia, Sjostrand:1993yb} is based on the Monte-Carlo implementation of the Lund string fragmentation function 
\beq
f(z, m_{T}) = \frac{1}{z} (1-z)^{a} \exp \left(-\frac{b\, m{_T}^2}{z} \right)
\eeq
where $m_{T}$ and $z$ denote the transverse mass and the light cone momentum fraction of the fragmenting hadron. The default parameters $a$ and $b$ are constrained by global analysis of p+p data. The hadronization scheme in PYTHIA also includes decays of hadron resonances and final-state hadronic interactions. 
A quantitative study of the effect of hadronization on initial-state parton level correlations using different schemes of fragmentation has been done in Ref~\cite{Esposito:2015yva}. The study indicates that the structure of the final azimuthal correlations are very much sensitive to the choice of fragmentation. For example the resonance decay and hadronic interactions can distort the initial-state correlations and introduce artificial final-state correlations in the produced hadrons. Such mechanism will complicate the interpretation of experimental data. In such a context it has been demonstrated that  hardronization effects in PYTHIA combined with the scheme of color reconnection produce effects that can mimic collective flow like pattern leading to a strong mass ordering of $\la p_T\ra$~\cite{Ortiz:2013yxa} in p+p collisions. In addition, the hadronic transport models that implements the Lund string fragmentation of PYTHIA have also shown to qualitatively reproduce the systematics of $v_2$ measured in p+Pb collisions~\cite{Zhou:2015iba}. Clearly more phenomenology in this direction will further improve our understanding. Meanwhile it is essential to implement a state-of-the art framework of hadronization to test different features of multi-particle correlations in the CGC at the level of hadrons; work in this direction is in progress~\cite{Schenke:2016lrs}. 
%


%
\begin{figure*}[t]
\includegraphics[width=0.45\textwidth]{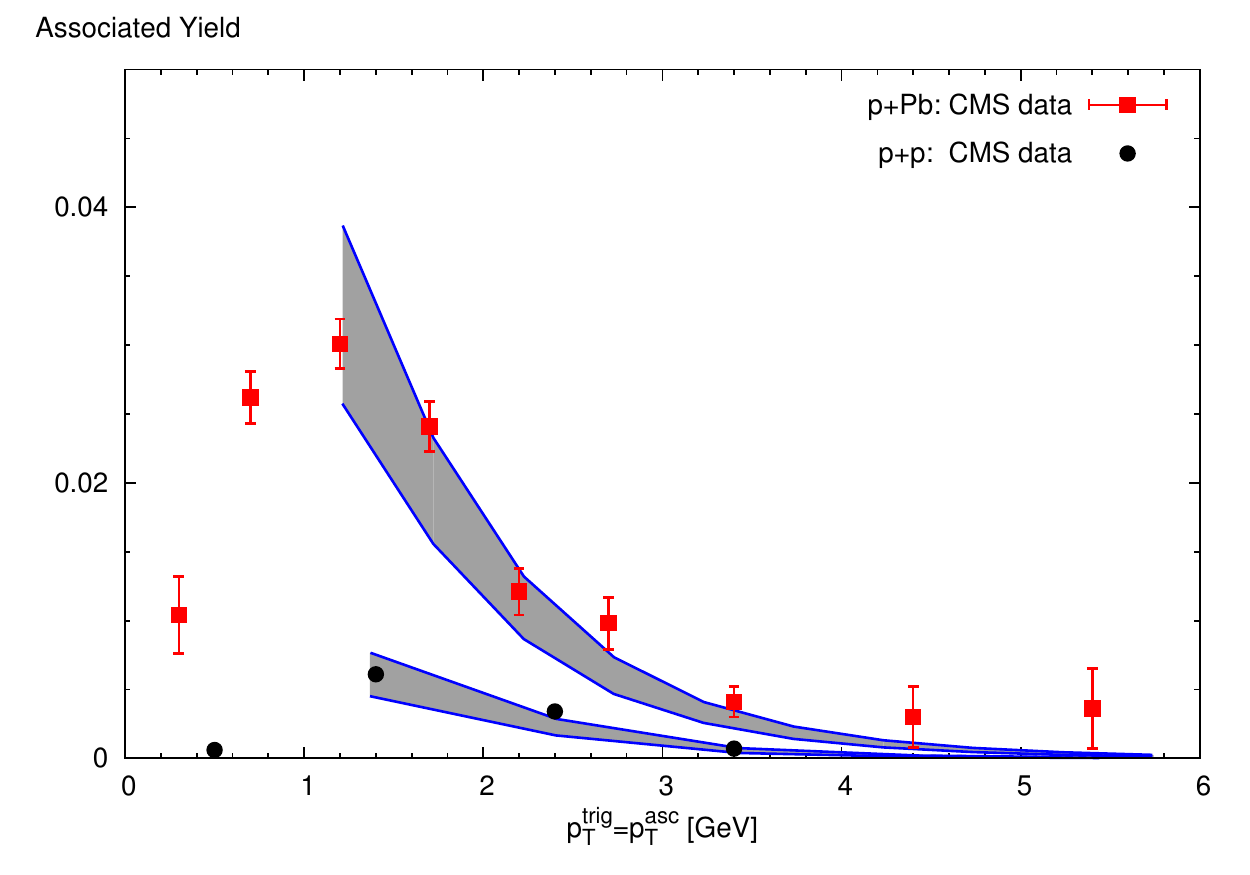}
\includegraphics[width=0.45\textwidth]{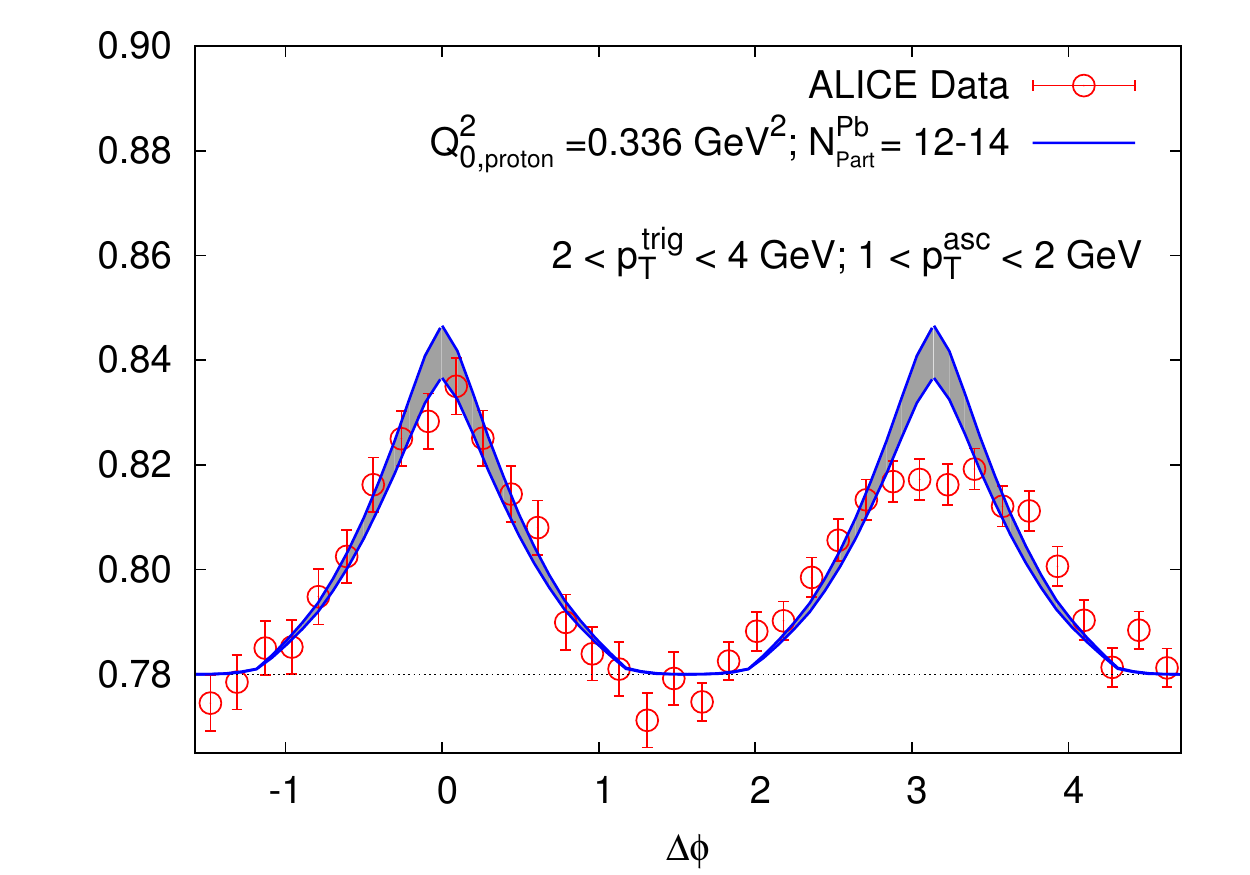}
\caption{\label{fig_pAridge} Figure showing the systematics of ridge-like correlations in p+p and p+Pb collisions captured in the glasma-graph framework taken from Ref\cite{Dusling:2013qoz}. The comparisons are made with the CMS data for p+p and p+Pb collisions for events having $N_{\rm trk}^{\rm offline} \ge 110$ from \cite{Khachatryan:2010gv, CMS:2012qk} and ALICE data for p+Pb collisions Ref.\cite{Abelev:2012ola}}
\end{figure*}

\section{Comparison with the latest experimental observations}
\label{seq::comparison}

\subsection{p+p collisions}
We begin with a comparison of model calculations with the latest data in $p+p$ collisions at LHC energies~\cite{Aad:2015gqa}, where the quantity of experimental interest is the rapidity and momentum integrated correlation function defined as 
\begin{align}
\label{eq:dihadron}
\frac{dN}{d\Delta \phi} =& 
\int\limits_{p_T^{\rm min}}^{p_T^{\rm max}} \frac{dp_T^2}{2} \int\limits_{q_T^{\rm min}}^{q_T^{\rm max}}\frac{ d q_T^2}{2}\;\int d\phi_p \int d\phi_q\; \delta\left(\phi_p-\phi_q-\Delta\phi\right) \nonumber\\
&\!\!\times
\frac{d^2N_{}^{\rm \sl corr.}}{d^2\pp d^2\qp d\eta_p d\eta_q}\left(\frac{p_{\textrm{T}}}{z_1},\frac{q_{\textrm{T}}}{z_2},\Delta\phi \right) 
\end{align}
which so far has only been computed in the perturbative CGC approach.
The quantity of experimental interest is the near-side yield $Y_{\rm int}$ defined as the ZYAM subtracted integrated associated yield per trigger given by  
\begin{equation}
\label{eq:zyam}
\textrm{Y}_{\rm int} = \frac{1}{\Ntrig}\int\limits_0^{\Delta\phi_{\rm min.}} \!\!\!\!
d\Delta\phi\left(\frac{dN}{d\Delta\phi}-\left.\frac{dN}{d\Delta\phi}\right|_{\Delta\phi_{\rm
min}}\right)\,. 
\end{equation}
Here $\Delta\phi=\Delta\phi_{\rm min.}$ corresponds to the minimum of the di-hadron correlation function (ZYAM) and $\Ntrig$ is the number of trigger particles. A quantitative comparison of the near-side yield in $p+p$ collisions between initial state calculations \cite{Dusling:2015rja} using the Glasma graph + BFKL approach and the experimental data at $7$ and $13$ TeV from the CMS collaboration~\cite{Khachatryan:2015lva} is shown in Fig.~\ref{fig_scaling}. One can see a good agreement between the perturbative calculations and the data at low multiplicities. At higher multiplicities one sees systematic deviations where the calculations over predict the data. An improved calculations in the framework of Classical Yang-Mills which consistently include multiple-scattering effects~\cite{Schenke:2015aqa} might lead to a better description of the data. 

The striking feature of the data model comparison shown in Fig.\ref{fig_scaling} is that the near-side yield is approximately energy independent. Such energy independent scaling is naturally explained within the framework of CGC \cite{Dusling:2015rja}. Origin of such scaling can be understood as follows. 
In the CGC framework a single scale $\qs^2$ determines both the single and double inclusive production in p+p collisions. The dependence on the center of mass energy in both multiplicity $\Nch$ and the near side yield $Y_{\rm int}$ enters only through $\qs^2$, i.e. 
\begin{equation}
\Nch (\qnot^2, \sqrt{s}) = \Nch (\qs^2) \, ,\,  Y_{\rm int} (\qnot^2, \sqrt{s}) = Y_{\rm int} (\qs^2), 
\end{equation} 
where $\qs^2 \equiv \qs^2(\qnot^2, \sqrt{s})$. Fixing the multiplicity $\Nch(\qs^2)$ therefore naturally fixes the value of $Y_{\rm int}(\qs^2)$ giving energy independent scaling of the ridge-like correlations. Such scaling provides strong indication of multi-particle production driven by a single semi-hard scale as captured in the CGC.

\subsection{p+A collisions}

\begin{figure*}[t]
\includegraphics[width=0.45\textwidth]{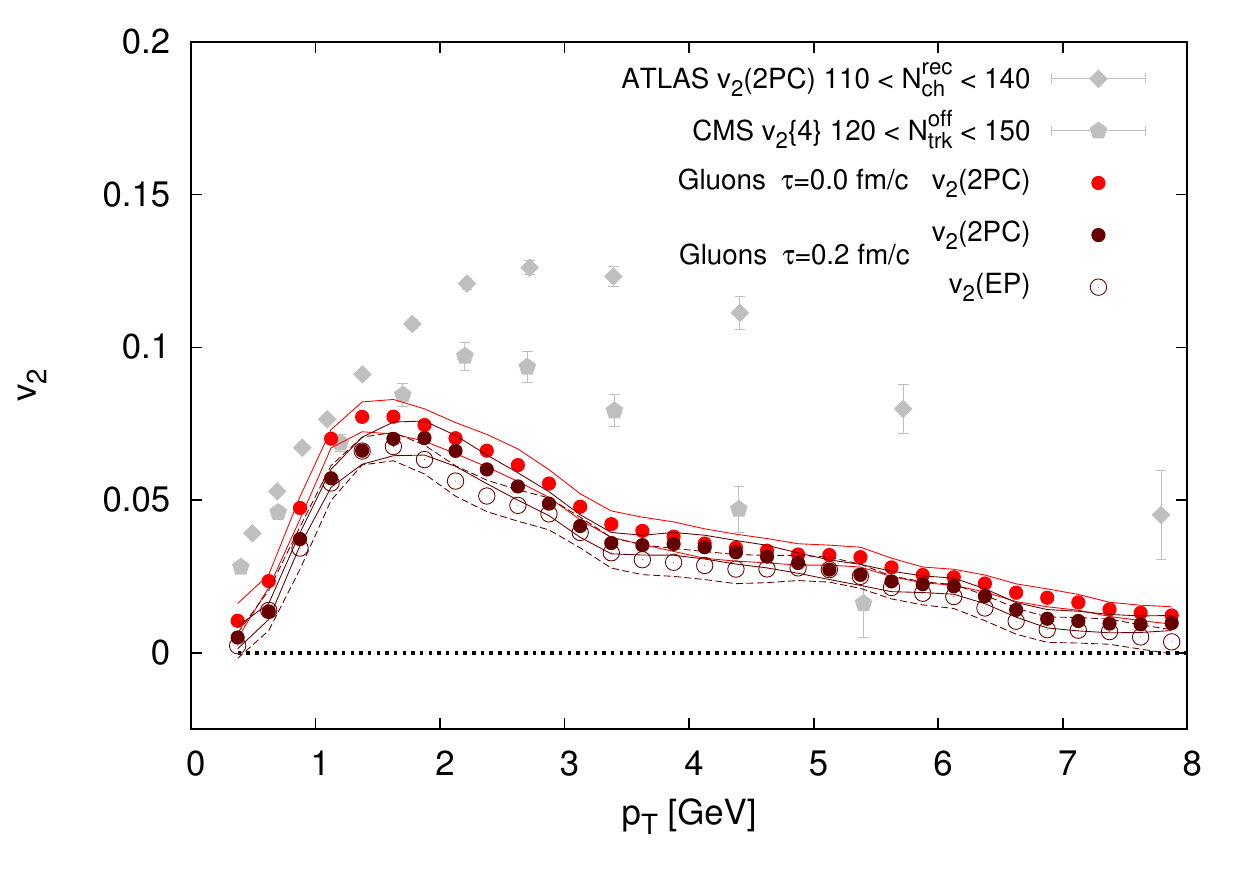}
\includegraphics[width=0.44\textwidth]{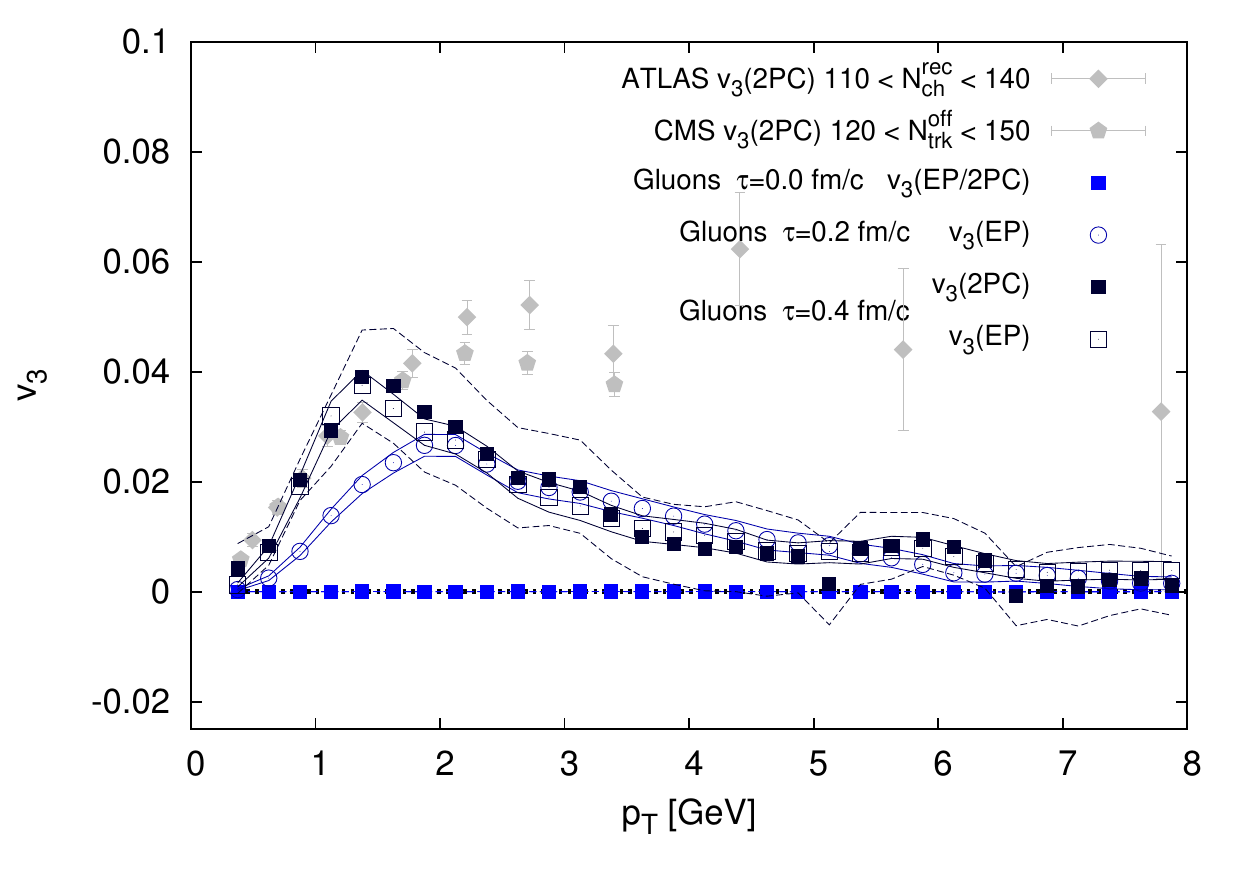}
\caption{\label{fig_cymvn} Right: Gluon $v_2(p_T)$ (left) and  $v_3(p_T)$ (right) in p+Pb collisions from Yang-Mills simulations, figure obtained from~\cite{Schenke:2015aqa} compared to data from ATLAS~\cite{Aad:2014lta} and CMS collaboration \cite{Chatrchyan:2013nka} for inclusive hadrons.}
\end{figure*}

A detailed quantitative estimation of the di-hadron correlations in p+Pb collisions at the LHC using the perturbative di-jet and glasma-graph framework has been performed in Ref~\cite{Dusling:2013qoz}. One of the interesting results from Ref~\cite{Dusling:2013qoz} shown in Fig.\ref{fig_pAridge} (left) indicates that the CGC-framework naturally explains the increase in the near side yield in p+Pb collisions as compared to p+p collisions for events with thesame multiplicity measured by the CMS collaboration. Results shown in Fig.\ref{fig_pAridge} (right) indicate that such framework also very well reproduces the peripheral subtracted correlation function $\left(1/N_{\rm trig}\right) dN/d\Delta\phi$ measured by the ALICE collaboration in p+Pb collisions. Aforementioned due to the limitations of the perturbative framework such estimations can be performed only for trigger and associated momentum $\pp, \qp \geq 1$ GeV. 

Recent simulations in the framework of classical Yang-Mills that go beyond such limitations of the perturbative regime have already opened the path for improved phenomenology. Even though at present these calculations do not yet include di-jet graphs\footnote{Note that also the interference contribution between Glasma graphs and jet graphs no longer vanishes beyond leading order.} and hadronization effects -- and thus do not allow for a direct comparison with experimental data -- simulations in this regard have lead to new insight into the correlations at the parton level concerning in particular the dynamics of the correlations during the very early stages. The results from the recent classical Yang-Mills simulations performed in p+Pb collisions ~\cite{Schenke:2015aqa} are shown in Fig.\ref{fig_cymvn}.  While gluons are produced with a significant momentum space anisotropy at $\tau=0^{+}$, initially the two-particle correlation function is  symmetric around $\Delta \phi=\pi/2$ and features only even harmonics in accordance with the perturbative result \cite{Schenke:2015aqa}. However, including the effects of the classical Yang-Mills evolution up to $\tau=0.4 \rm{fm}/c$ leads to the build up of a sizable $v_{3}$ on the parton level, while the initial state $v_{2}$ remains intact  \cite{Schenke:2015aqa}. These results indicate that at least at the gluon level the non-perturbative dynamics of the glasma fields gives rise to large values of both $v_2$ and $v_3$ at time scales $\tau\sim 1/Q_s$ after the collision that are comparable to what is seen in the data. Interestingly sizable $v_2$ and $v_3$ in this framework also extends to large transverse momentum probably beyond the regime of the applicability of hydro in p+Pb collisions~\cite{Niemi:2014wta}.  

While a systematic comparison of the initial state calculations to experimental data in $p+Pb$ collisions yields a similar level of quantitative agreement of two-particle correlations for momenta $p_T > 1~\rm{GeV}$ \cite{Dusling:2012wy,Dusling:2013qoz}, there has been enormous progress on the experimental side to identify additional signatures of collective motion which could be indicative of the onset of significant final state effects. Even though some observables, such as e.g. mass ordering properties observed in correlations between identified hadrons \cite{Abelev:2013wsa} are not necessarily sensitive to the origin of the correlations, more promising directions including e.g. correlations between more than two particles have also been explored \cite{Khachatryan:2015waa,Abelev:2014mda}. One of the most striking observations in this regard is the sign change of the four-particle cumulant $c_{2}\{4\}$ observed around $N_{ch}(|\eta_{lab}|<1) \sim 60$ by ALICE \cite{Abelev:2014mda}. However, most of the recent measurements have focused on low $p_T$ observables where the perturbative calculations \cite{Dusling:2012iga,Dusling:2012cg,Dusling:2012wy,Dusling:2013qoz,Dusling:2015rja} outlined in Sec.~\ref{seq_pertub} do not necessarily apply, hence complicating the comparison between theory and experiment.

Nevertheless, there have been first attempts to extend the theoretical framework to understand whether certain features of the low $p_T$ data such as the sign change of $c_{2}\{4\}$ can also be explained from initial state effects. Based on the dilute-dense approximation in Eq.~(\ref{eq:E-dipole}) first attempts have been made to study initial-state correlations of more than two particles. Generalizing the dilute-dense formalism to $n$-particle scattering, it was pointed out in \cite{Dumitru:2014yza,Skokov:2014tka,Lappi:2015vta} that the four particle cumulant  $c_{2}\{4\}$ is sensitive to non-Gaussian correlations of the color-electric fields inside the target. While the lowest order perturbative contribution to $c_{2}\{4\}$ was found to be positive, it was argued \cite{Dumitru:2014yza} that non-Gaussian correlations of the domains of color-electric fields may explain the experimentally observed sign change of $c_{2}\{4\}$  as a function of multiplicity. Even though a dynamical explanation for the existence of sizable non-Gaussian correlations is still lacking, this topic is an active subject of further investigations.  However, one should also caution that in contrast to the two particle cumulant $c_{2}\{2,|\Delta \eta >2\}$, where short-range correlations are suppressed by introducing a large rapidity gap, the four-particle cumulant $c_{2}\{4\}$ also receives contributions from short-range correlations and interference diagrams. While a complete theoretical calculation of higher order $n$-particle correlations is desirable, further progress is needed to tackle this problem.

Similarly, preliminary results from event-by-event simulations in classical Yang-Mills theory a la \cite{Schenke:2015aqa} also suggest a negative sign of $c_{2}\{4\}$ when all particles have low momenta, while at high $p_T$ the four particle cumulant is always positive \cite{soeren2015qm}. While the first results are interesting, further theoretical progress is needed to decide unambiguously whether the sign change in $c_2\{4\}$ can be explained in terms of initial state and early-time effects. It would also be interesting to extend the experimental measurements of higher cumulants towards higher momenta to achieve convergence on this issue.

\section{Summary}

Experimental observations of collective correlations in small systems challenge our current 
understanding of the space-time evolution of high-energy collisions. While at low multiplicities 
one expects dominance of initial state effects, it is also expected that at sufficiently
high multiplicities initial state correlations are destroyed due to final state interactions and the hydrodynamic response to
the initial state geometry provides the dominant source of correlations.
However, it is theoretically challenging to predict the transition from initial state to final state
dominated dynamics pointing to the importance of developing a unified theoretical framework 
where both effects are consistently taken into account. 

In this brief review we specifically outlined recent developments in the approach based on initial state dynamics. While calculations based purely on initial state correlations are able to quantitatively describe various features of the experimental data in p + p and p+Pb collisions up to the highest multiplicity windows, a simultaneous description of the low and high $p_T$ observables over a wide range of multiplicity remains challenging within any single theoretical framework. Of course, several outstanding issues remain on the theory side and the development of a unified framework that combines 1) dynamics of multiple-soft interactions with 2) a first principle calculation of jet production and 3) a state-of-the-art fragmentation scheme is essential for a wide range of phenomenological applications. So far experimental efforts have focused on soft observables such as e.g. the $v_n$s, however complimentary information on the dynamics can be obtained by considering higher-momentum probes. While in heavy-ion collisions, the observation of strong jet-quenching phenomena provides an important indication for the formation of a strongly interacting Quark-Gluon plasma, no such features have been reported so far in small systems.  Even though such an analysis is experimentally challenging, it would be important to establish to what extent mini-jets and actual jets are modified in high-multiplicity events to further constrain the relative importance of initial state and final state effects in small systems.
 
 
 \section*{Acknowledgement}
We thank Kevin Dusling, Ulrich Heinz, Tuomas Lappi, Wei Li, Derek Teaney and Raju Venugopalan for important discussions. We thank Bjoern Schenke for careful reading of the manuscript and helpful comments and suggestions. The authors are supported under Department of Energy contract number Contract No. DE-SC0012704. S. Schlichting acknowledges support under DOE Grant no. DE-FG02-97ER41014.

\bibliographystyle{apsrev4-1}
\bibliography{references}

\end{document}